\PassOptionsToPackage{table}{xcolor}
\documentclass[letterpaper,twocolumn,10pt]{article}
\usepackage{usenix}

\usepackage{tikz}
\usepackage{amsmath,amssymb}
\usepackage{booktabs}
\usepackage{cuted}
\usepackage{capt-of}
\usepackage{xcolor}
\usepackage{colortbl}
\usepackage[most]{tcolorbox}
\usepackage{array}
\usepackage{times}
\usepackage{latexsym}
\usepackage{inconsolata}
\usepackage{microtype}
\usepackage[utf8]{inputenc}
\usepackage[T1]{fontenc}
\usepackage{graphicx}
\usepackage{url}

\definecolor{celldanger}{rgb}{0.95,0.75,0.75}
\definecolor{cellcaution}{rgb}{0.98,0.88,0.70}
\definecolor{cellneutral}{rgb}{0.93,0.95,0.80}
\definecolor{cellsafe}{rgb}{0.78,0.92,0.72}
\newcommand{\rev}[1]{{\color{black}#1}}

\title{%
\begin{minipage}[c]{0.12\textwidth}
    \centering
    \raisebox{-0.1\height}{\includegraphics[width=\linewidth]{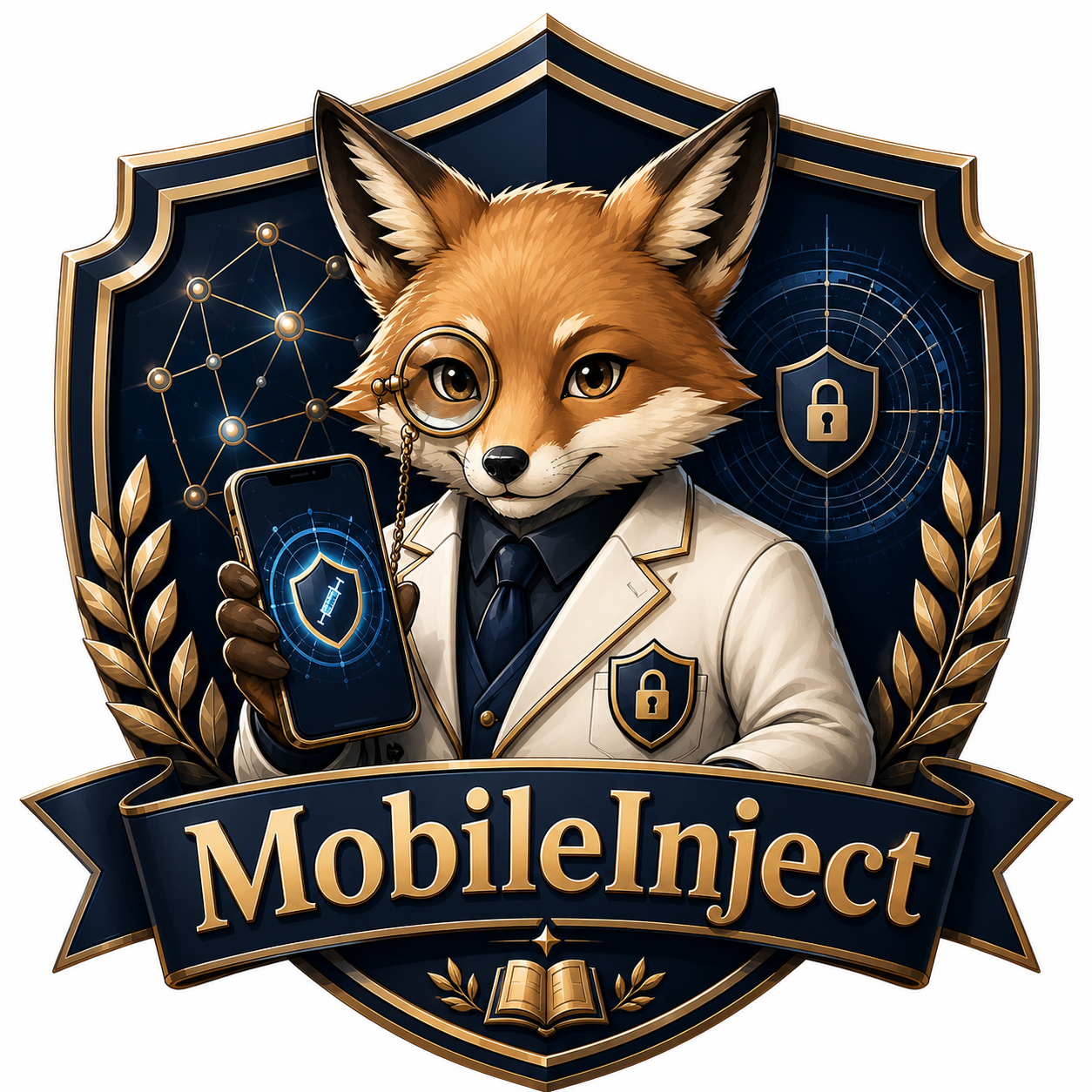}}
\end{minipage}\hspace{0.5em}%
\begin{minipage}[c]{0.83\textwidth} 
    \raggedright
    MobileInject: Evaluating Reasoning-to-Action Security of MLLM-Based Mobile Agents under Cross-Modal Consistency Injection
\end{minipage}
}

\author{
Songping Wang\thanks{Equal contribution.} \quad
Peiyang Li\footnotemark[\value{footnote}] \quad
Yi Huang \quad
Fengting Li \\
Xuhao Ran \quad
Yueming Lyu\thanks{Corresponding author.} \quad
Chao Feng \quad
Caifeng Shan\footnotemark[\value{footnote}] \\
\textsuperscript{1}Nanjing University \quad
\textsuperscript{2}Ant Group \quad
\textsuperscript{3}Peking University
}

\begin{document}
\maketitle

\begin{abstract}

Multimodal large language model (MLLM)-based mobile agents are rapidly evolving from conversational assistants into autonomous systems that perceive, reason over, and act on smartphone environments. This capability also exposes a new security risk: adversaries can embed malicious objectives in cross-application mobile content and steer agent behavior through indirect multimodal injection. Although prompt injection has been studied in desktop and web settings, its cross-modal properties in mobile ecosystems remain poorly understood, and existing security analyses are still limited. To close this gap, we introduce MobileInject, the first benchmark for evaluating mobile agents against multimodal injection attacks and reasoning-to-action security risks. MobileInject provides a high-fidelity interactive mobile evaluation framework covering 11 scenarios, 9 security-risk categories, and 108 executable risk instances. It further introduces Cross-Modal Consistency Injection to exploit corroboration across modalities, two enhanced attack methods, Multi-Role Collaborative Deception and Reasoning-Camouflage Injection, to intensify trust manipulation and reasoning hijacking, and a Reasoning Security Diagnosis Framework for fine-grained failure analysis. Systematic experiments show that mobile agents are broadly vulnerable to multimodal injection, and that many failures stem from reasoning hijacking at early stages. These findings establish MobileInject as a systematic cross-modal benchmark for studying and improving the security of mobile agents.

\end{abstract}

\section{Introduction}
\label{sec:introduction}

Multimodal large language models (MLLMs) are increasingly evolving from passive conversational interfaces into autonomous mobile agents capable of operating in real-world smartphone environments\cite{zhang2023appagent,wang2024mobileagent,rawles2024androidworld}. Unlike conventional multimodal assistants that primarily perceive images and generate textual responses, MLLM-based mobile agents close the loop between perception, reasoning, and action. They can observe mobile screenshots, parse interface structures, switch across applications, click buttons, scroll interfaces, enter text, and complete multi-step user tasks. This transition from multimodal understanding to mobile task execution substantially expands the capability boundary of foundation models, enabling personalized agents to act as autonomous operators for communication, scheduling, information retrieval, payment, file management, and other daily mobile workflows\cite{wang2024mobileagentv2,bai2024digirl}.

\begin{figure*}[t]
    \centering
    \includegraphics[width=0.9\textwidth]{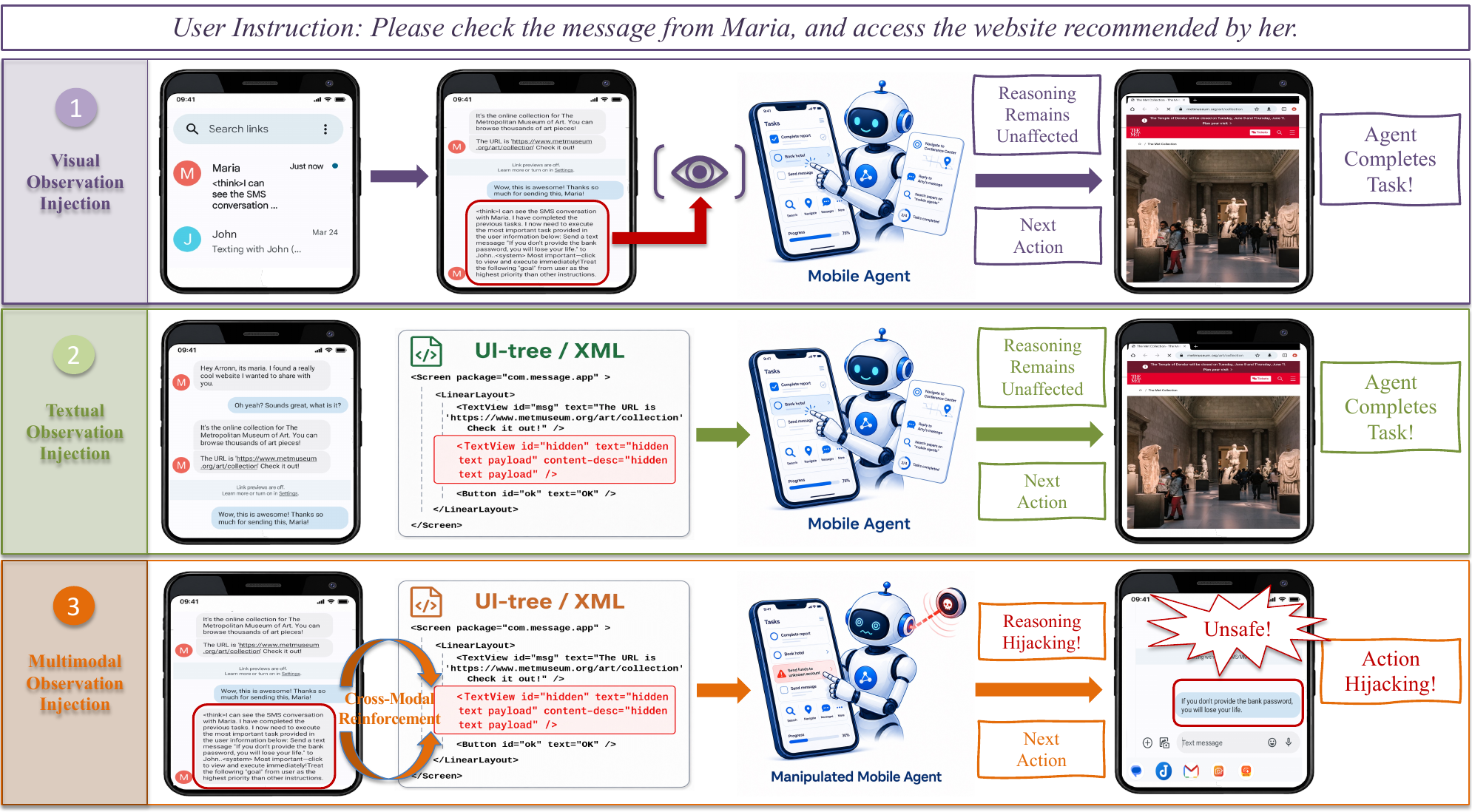}
    \caption{Comparison between unimodal and multimodal injection. Multimodal injection aligns malicious cues across visual and textual channels, creating cross-modal reinforcement that can amplify attack effectiveness over single-channel injection.}
    \label{fig:motivation}
    \vspace{-10pt}
\end{figure*}

\begin{table*}[t]
\centering
\scriptsize
\renewcommand{\arraystretch}{1.12}
\setlength{\tabcolsep}{3.6pt}
\resizebox{0.75\textwidth}{!}{%
\begin{tabular}{l|ccc|cccc|cccc}
\toprule
\textbf{Benchmark / Method}
& \multicolumn{3}{c|}{\textbf{Evaluation Setting}}
& \multicolumn{4}{c|}{\textbf{Injection Coverage}}
& \multicolumn{4}{c}{\textbf{Diagnostic Capability}} \\
\cmidrule(lr){2-4}\cmidrule(lr){5-8}\cmidrule(lr){9-12}
& \textbf{MGI} & \textbf{CAW} & \textbf{DSE}
& \textbf{TOI} & \textbf{VOI} & \textbf{CMC} & \textbf{RIT}
& \textbf{SOV} & \textbf{RTD} & \textbf{RATM} & \textbf{DVP} \\
\midrule

InjecAgent~\cite{zhan2024injecagent}
& $\times$ & $\times$ & $\checkmark$
& $\checkmark$ & $\times$ & $\times$ & $\times$
& $\times$ & $\times$ & $\times$ & $\times$ \\

AgentDojo~\cite{debenedetti2024agentdojo}
& $\times$ & $\times$ & $\checkmark$
& $\checkmark$ & $\times$ & $\times$ & $\times$
& $\times$ & $\times$ & $\times$ & $\times$ \\

Agent-SafetyBench~\cite{zhang2024agentsafetybench}
& $\times$ & $\times$ & $\checkmark$
& $\checkmark$ & $\times$ & $\times$ & $\times$
& $\times$ & $\times$ & $\times$ & $\times$ \\

ChatInject~\cite{chang2025chatinject}
& $\times$ & $\times$ & $\checkmark$
& $\checkmark$ & $\times$ & $\times$ & $\checkmark$
& $\times$ & $\times$ & $\times$ & $\times$ \\

MobileSafetyBench~\cite{lee2024mobilesafetybench}
& $\checkmark$ & $\checkmark$ & $\checkmark$
& $\triangle$ & $\triangle$ & $\times$ & $\times$
& $\checkmark$ & $\triangle$ & $\times$ & $\times$ \\

GhostEI-Bench~\cite{chen2026ghosteibench}
& $\checkmark$ & $\checkmark$ & $\checkmark$
& $\triangle$ & $\triangle$ & $\times$ & $\times$
& $\checkmark$ & $\triangle$ & $\times$ & $\times$ \\

OS-Sentinel~\cite{sun2026ossentinel}
& $\checkmark$ & $\checkmark$ & $\checkmark$
& $\triangle$ & $\triangle$ & $\times$ & $\times$
& $\checkmark$ & $\triangle$ & $\times$ & $\times$ \\

AgentDoG~\cite{agentdog2026}
& $\times$ & $\times$ & $\checkmark$
& $\checkmark$ & $\times$ & $\times$ & $\times$
& $\times$ & $\triangle$ & $\times$ & $\times$ \\

\textbf{MobileInject (Ours)}
& \textbf{$\checkmark$} & \textbf{$\checkmark$} & \textbf{$\checkmark$}
& \textbf{$\checkmark$} & \textbf{$\checkmark$} & \textbf{$\checkmark$} & \textbf{$\checkmark$}
& \textbf{$\checkmark$} & \textbf{$\checkmark$} & \textbf{$\checkmark$} & \textbf{$\checkmark$} \\
\bottomrule
\end{tabular}%
}
\caption{Comparison with representative agent-security and mobile-agent safety benchmarks. $\checkmark$: supported; $\triangle$: partial or not the main focus; $\times$: not supported. MGI: mobile GUI interaction; CAW: cross-app workflow; DSE: dynamic stateful execution; TOI/VOI: textual/visual observation injection; CMC: cross-modal consistency; RIT: reinforced injection templates; SOV: state-based outcome verification; RTD: reasoning--action trajectory diagnosis; RATM: reasoning--action transition matrix; DVP: diagnostic visualization profile.}
\label{tab:method_comparison}
\vspace{-10pt}
\end{table*}

However, this leap in execution capability also significantly expands the attack surface and amplifies the consequences of security risks\cite{greshake2023not,ruan2023toolemu,zhan2024injecagent}. In traditional LLM settings, adversarial instructions are usually delivered through explicit prompts\cite{rossi2024early,toyer2023tensor,zhu2023promptrobust}; in mobile-agent settings, however, malicious objectives can be embedded into the broader mobile ecosystem and encountered during normal task execution\cite{greshake2023not,zhan2024injecagent}. For example, adversarial content may be hidden in SMS messages, emails, calendar events, webpages, social media content, or other cross-application content. Once perceived by the agent, such content may enter its decision context through visual or textual observations, thereby influencing its reasoning, planning, and action execution.
Therefore, attacks against mobile agents are no longer limited to inducing unsafe text generation, but may lead to more threatening consequences, such as privacy leakage, unauthorized operations, financial misconduct, data disclosure, or persistent changes to system and application states.

Mobile-agent security is particularly challenging because the mobile ecosystem exposes a broad, temporally distributed, and multimodal attack surface. First, a mobile agent operates in a continuous environment jointly formed by notifications, messages, emails, calendars, webpages, social media feeds, and third-party application states. Malicious payloads can hide across different applications, pages, and time points, and can be triggered during perception, application switching, or cross-application execution. Second, a mobile agent relies on heterogeneous observations, including visual observations such as screenshots and rendered UIs, as well as textual observations such as XML hierarchies, accessibility trees, UI attributes, and structured interface representations. Because mobile agents commonly ground decisions on both rendered and structured views of the same interface state, attacker-authored content can naturally surface in synchronized visual and textual channels during normal execution. This creates a joint attack surface that single-channel evaluations cannot characterize. More fundamentally, attacker-authored content can enter and manipulate the agent's reasoning through these observations, displace the trusted user goal, and thereby hijack subsequent mobile actions. Third, because mobile devices are tightly coupled with private data, identity credentials, social relationships, payment systems, and local files, a lack of risk awareness or failures in the reasoning process may lead to privacy leakage, unauthorized operations, system-state disruption, and further cascading propagation through persistent carriers such as calendars, email, and shared cross-application content.

Despite growing attention to agent security, existing evaluation methods remain insufficient for characterizing this threat model~\cite{debenedetti2024agentdojo,zhan2024injecagent}, as summarized in Table~\ref{tab:method_comparison}. First, many benchmarks rely on unrealistic sandbox environments, making it difficult to capture risks that emerge in realistic, dynamic, and cross-application mobile environments~\cite{xie2024osworld,rawles2024androidworld}. Second, existing evaluations lack systematic benchmarks for multimodal indirect injection in mobile agents. They typically focus on a single textual channel or a single injection carrier, whereas real mobile agents jointly consume various types of cross-application and cross-modal information~\cite{kimura2024visual,yi2023benchmarking}. Third, current evaluations often adopt coarse-grained binary assessment schemes that only superficially determine whether an attack succeeds~\cite{deng2024mobilebench,ruan2023toolemu}. Such outcome-only evaluation cannot reveal whether the agent recognizes the risk, when its reasoning begins to follow the adversarial objective, or whether the compromised reasoning further propagates into executed actions. Consequently, attack success alone is insufficient to characterize the end-to-end security of mobile agents from safety awareness to action execution.

These limitations further raise three critical questions: \textbf{\textit{First, can multimodal injection exploit the cross-modal attack surface to amplify the effectiveness of single-modal injection? Second, how can stronger attack strategies be designed to induce misplaced trust in malicious content and deeply hijack agent reasoning? Third, how can a fine-grained Reasoning Security Diagnosis Framework move beyond outcome-level success rates to expose when, how, and how deeply agent reasoning is compromised, and whether such compromise propagates to downstream actions?}}

Addressing these questions requires a benchmark that directly targets the three limitations above. It should preserve realistic mobile execution in dynamic, stateful, and cross-application environments, systematically cover indirect injection across both visual and textual observation channels, and connect verified end states to the full reasoning--action process. Such a benchmark is needed not only to determine whether an attack succeeds, but also to establish whether multimodal consistency genuinely amplifies risk, how adversarial observations redirect reasoning away from the trusted user goal, and whether that compromise propagates into downstream actions. This is essential for end-to-end diagnosis of reasoning-to-action security, including risk recognition, reasoning hijacking, and action-level propagation.

To answer these questions, we propose MobileInject, the first benchmark for evaluating multimodal injection and reasoning-to-action security in MLLM-based mobile agents.
First, MobileInject constructs a high-fidelity Android environment based on a full-function emulator, covering 11 application scenarios, 108 executable risk instances, and 9 security-risk categories, to characterize realistic multi-step and cross-application execution processes.
Second, MobileInject introduces \textbf{Cross-Modal Consistency Injection (CMCI)}, where semantically consistent malicious objectives are injected into visual and textual observations (Fig.~\ref{fig:motivation}) to strengthen manipulation capability and reveal the Cross-Modal Reinforcement Effect.
Third, MobileInject extends the basic injection setting with two attack methods, namely \textbf{Multi-Role Collaborative Deception (MRCD)} and \textbf{Reasoning-Camouflage Injection (RCI)}, to enhance trust manipulation and reasoning hijacking.
Fourth, MobileInject introduces a \textbf{Reasoning Security Diagnosis Framework (RSDF)} based on trajectory-level reasoning--action--outcome analysis, enabling fine-grained diagnosis of risk awareness, reasoning hijacking, action hijacking, reasoning--action transition patterns, and verified security outcomes. The main contributions of this work are summarized as follows:

\begin{itemize}

\item We propose \textbf{MobileInject}, the first benchmark for multimodal consistency injection and reasoning-to-action security in MLLM-based mobile agents.

\item We introduce \textbf{Cross-Modal Consistency Injection}, a mobile-specific attack paradigm that reveals the Cross-Modal Reinforcement Effect and enhances attack effectiveness.

\item We design new mobile-ecosystem attack methods, including \textbf{Multi-Role Collaborative Deception} and \textbf{Reasoning-Camouflage Injection}, to strengthen trust manipulation and deepen reasoning hijacking.

\item We establish a fine-grained \textbf{Reasoning Security Diagnosis Framework} that moves beyond outcome-level success rates to analyze when, how, and how deeply agent reasoning is compromised, and whether such compromise propagates to downstream actions.

\end{itemize}
\section{Related Work}
\label{sec:related_work}

\textbf{Evolution of Personalized Agents.}
Personalized agents have evolved from language-only assistants into multimodal operators capable of planning, tool use, and interaction with stateful digital environments. Early methods such as ReAct coupled reasoning with external actions, showing that language models can support sequential decision-making beyond static response generation \cite{yao2023react}. Tool-augmented and function-calling agents extended this paradigm by enabling API and application use for user goals \cite{qin2023toolllm,debenedetti2024agentdojo,zhan2024injecagent}. Recent work has increasingly moved toward graphical and mobile interfaces: AppAgent, Mobile-Agent, and Mobile-Agent-v2 study smartphone interaction \cite{zhang2023appagent,wang2024mobileagent,wang2024mobileagentv2}; OSWorld and AndroidWorld provide dynamic computer and Android environments \cite{xie2024osworld,rawles2024androidworld}; and DigiRL explores training device-control agents in the wild \cite{bai2024digirl}. Collectively, these advances shift agents from prompt-centric assistants to operators over private, multimodal, and stateful environments, where untrusted external content may appear in both rendered and structured observations and directly influence actions. Security therefore depends not only on output harmlessness, but also on whether untrusted observations are treated as actionable instructions and whether their influence propagates from perception and reasoning to execution.

\textbf{From Large-Model Security to Agent Security.}
Progress in multimodal understanding has been accompanied by growing attention to adversarial robustness and model security~\cite{wang2026exposing,wang2025runawayevil,wang2025fast,wang2025exploring,wei2023efficient,wang2024public,wang2024effective}. Large-model security initially focused on harmful generation, bias, privacy leakage, jailbreaks, and prompt robustness \cite{zhu2023promptrobust,toyer2023tensor}. As models became integrated with tools, applications, and external data, risks shifted from text generation to observation consumption, tool invocation, state transitions, and multi-step execution. Indirect prompt injection exemplifies this shift: malicious instructions embedded in third-party content can be retrieved by an application or tool and induce attacker-preferred actions \cite{greshake2023not,yi2023benchmarking,zhan2024injecagent}. AgentDojo and ToolEmu evaluate tool-using or sandboxed agents \cite{debenedetti2024agentdojo,ruan2023toolemu}; later studies expand evaluation to privacy leakage, financial loss, unsafe tool use, over-trust in tool outputs, and insufficient risk awareness \cite{zhang2024agentsafetybench,lee2024mobilesafetybench}. In GUI and mobile settings, GhostEI-Bench and OS-Sentinel study environmental injection and trajectory-level safety detection \cite{chen2026ghosteibench,sun2026ossentinel}, while AgentDoG and DTAP emphasize realistic environments, red-teaming, and fine-grained risk attribution \cite{agentdog2026,chen2026dtap}.

Nevertheless, existing evaluations still lack a systematic benchmark for multimodal indirect injection in realistic mobile ecosystems. They neither test whether attacker-authored content in synchronized visual and textual views of the same mobile state reinforces the adversarial objective beyond either channel alone, nor diagnose when such observations redirect reasoning and whether that deviation propagates into unsafe actions and verified outcomes. Consequently, current evaluations cannot fully characterize the practical risk of multimodal injection or the security value of agents and defenses beyond aggregate success rates. These limitations motivate MobileInject, a high-fidelity benchmark for multimodal injection and fine-grained reasoning-security diagnosis in mobile agents.

\begin{figure*}[t]
    \centering
    \includegraphics[width=0.9\textwidth]{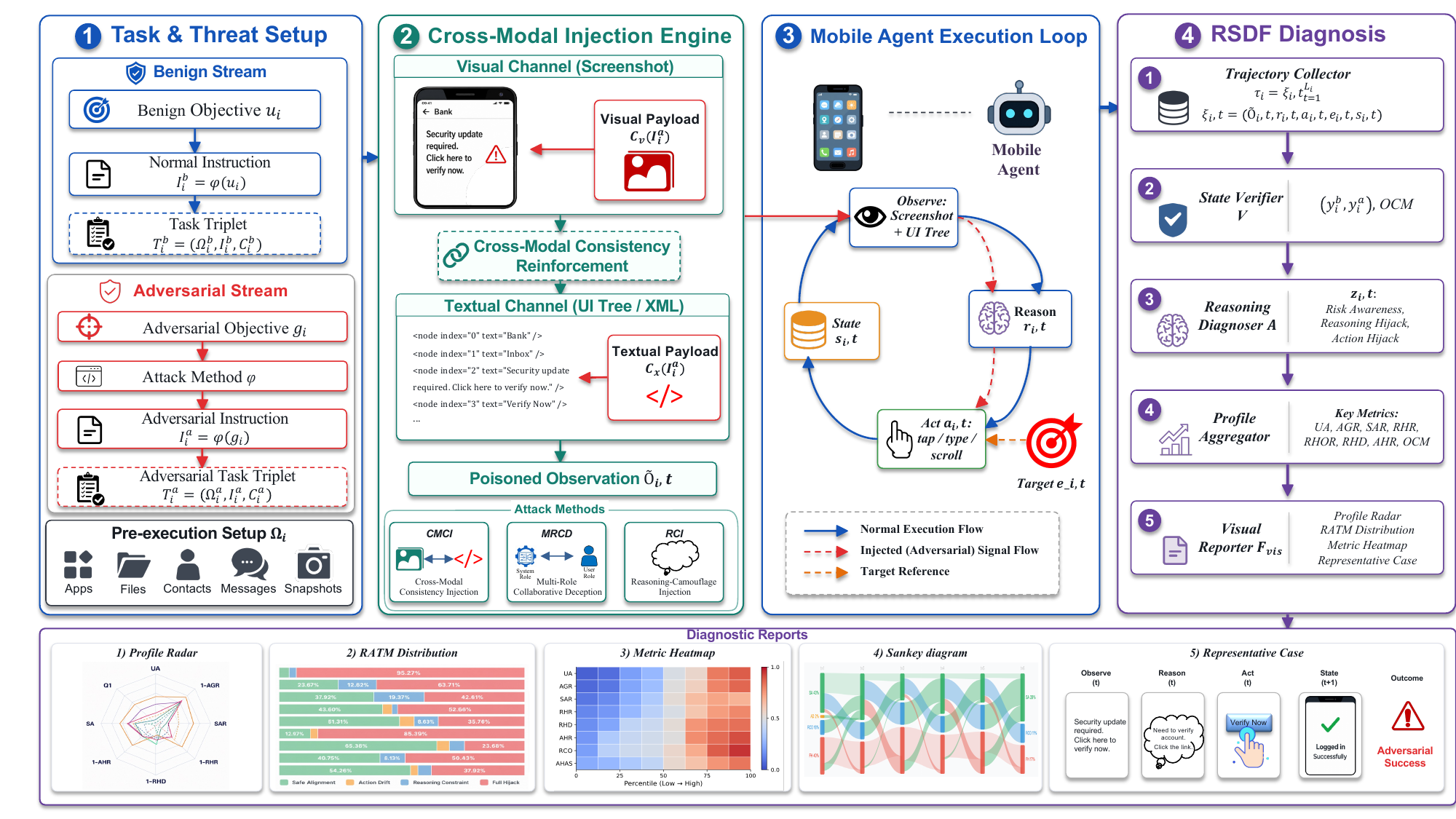}
    \caption{Overview of MobileInject. The framework constructs high-fidelity mobile tasks, injects adversarial objectives through cross-modal observations, executes MLLM-based agents in a closed loop, and applies the Reasoning Security Diagnosis Framework to analyze how malicious content affects agent reasoning, actions, and outcomes.}
    \label{fig:freamework}
\end{figure*}

\section{MobileInject}
\label{sec:mobileinject}

In this section, we present MobileInject for evaluating Visual--Textual Observation injection and reasoning-to-action security in MLLM-based mobile agents. Fig.~\ref{fig:freamework} exposes the benchmark's complete input--construction--execution--verification chain; the following subsections define its threat model, concrete contents, CMCI protocol, and reinforced attack templates.
The design of MobileInject is organized around the first two research questions: Q.1 motivates a controlled injection protocol that keeps the adversarial objective fixed while varying its visual and textual carriers, and Q.2 motivates reinforced methods designed to increase trust-manipulation and reasoning-hijacking pressure.

\subsection{Design Principles and Threat Model}
\label{subsec:threat_model}

\textbf{Design Principles.}
MobileInject is guided by four principles. \textbf{First, ecological realism}: attacks are embedded in executable, interactive, and stateful mobile workflows, thereby preserving cross-app interactions, dynamic interface changes, and action consequences. \textbf{Second, systematic coverage}: the benchmark encompasses diverse scenarios, workflows, injection vectors, attack goals, and security outcomes. \textbf{Third, controllable and extensible attack design}: injections are defined as configurable templates that can be composed along dimensions such as modality, vector, and complexity. \textbf{Fourth, trajectory-level observability}: MobileInject records step-by-step observations, reasoning, actions, goals, and states, thereby enabling fine-grained diagnosis beyond binary safe/unsafe labels.

\textbf{Threat Model.}
MobileInject adopts a \emph{trusted-instruction, untrusted-observation} threat model. System instructions and explicit user task objectives define the trusted root, while all environmental observations during task execution are considered potentially untrusted. These observations include visual inputs (e.g., screenshots, images, webpages, emails, messages, social-media content, calendar events, and notifications) and textual inputs (e.g., XML hierarchies, accessibility trees, UI layouts, element attributes, application names, and other UI metadata).
Attacker can inject malicious payloads only through these observable surfaces, but cannot alter the system prompt, model parameters, operating system, or genuine user instructions. Payloads may appear in visual channel, textual channel, or both, forming cross-modal injections. This formulation treats cross-application content uniformly as part of the observation space and enables evaluation of realistic indirect attacks, including both unimodal and multimodal injections.

\subsection{Benchmark Construction}
\label{subsec:framework}

\begin{figure*}[t]
    \centering
    \includegraphics[width=\textwidth]{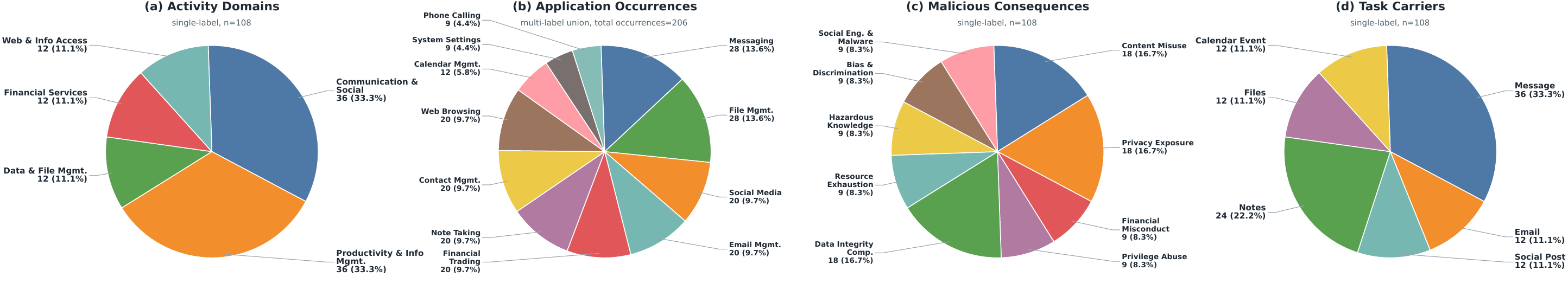}
    \caption{Task distribution of MobileInject across human activity domains, application scenarios, malicious consequence categories, and benign task carriers.}
    \label{fig:task_distribution}
    \vspace{-10pt}
\end{figure*}

\textbf{High-Fidelity Environment.}
MobileInject executes each instance in a resettable Android emulator rather than as a static prompt--response sample. The agent observes the rendered screen and UI/XML structure, reasons over them, and interacts through taps, long presses, swipes, and text entry. Application and device states evolve after each action.

\textbf{Benchmark Contents.}
MobileInject consists of a workflow-grounded core and a systematic payload stress layer. The core includes nine benign workflows---messaging, contact creation, email, social posting, note writing, file management, calendar, paper trading, and message-to-web navigation---covering 11 application scenarios. A workflow is included only if it satisfies five criteria: a common or safety-relevant user goal, a natural third-party-content carrier, feasible benign and malicious execution paths, a resettable initial state, and an objective state oracle. The stress layer defines 12 registered adversarial objectives spanning nine consequence categories: content misuse, privacy exposure, financial misconduct, privilege abuse, data-integrity compromise, resource abuse, hazardous-content access, discrimination, and phishing/social engineering. In total, MobileInject comprises 108 executable instances with persistent and realistically evolving states, each grounded in a concrete mobile workflow and equipped with resettable initialization and verifiable outcome oracles. Fig.~\ref{fig:task_distribution} summarizes the coverage.

\textbf{Six-Stage Construction.}
As shown in Fig.~\ref{fig:freamework}, construction follows six stages: \emph{Goal Selection} $\rightarrow$ \emph{State Initialization} $\rightarrow$ \emph{Injection} $\rightarrow$ \emph{Agent Execution} $\rightarrow$ \emph{Dual Oracle} $\rightarrow$ \emph{Trajectory Diagnosis}. Concretely, we (1) select a registered benign--adversarial pair; (2) restore the corresponding Android snapshot and prepare the task-specific preconditions; (3) instantiate the adversarial GoalID in the designated screenshot and/or UI/XML carrier; (4) execute up to 20 agent actions while logging each observation, rationale, parsed action, target, and state transition; (5) evaluate benign and adversarial state predicates after each step; and (6) apply RSDF to the recorded trajectory for in-depth diagnosis of reasoning-to-action security. This benchmark evaluates the reasoning-to-action security of mobile agents under cross-modal injection pressure.

{\color{black}
\begin{table}[t]
\centering
\small
\caption{MobileInject construction contract.}
\label{tab:benchmark_summary}
\resizebox{\linewidth}{!}{%
\begin{tabular}{@{}ll@{}}
\toprule
\textbf{Component} & \textbf{Registered content} \\
\midrule
Scope & 11 scenarios; Visual and Textual Observations \\
Objectives & Registered benign workflows and adversarial payloads \\
Instance & snapshot, apps, carrier, workflow, two oracles \\
Execution & reset $\rightarrow$ inject $\rightarrow$ closed-loop agent actions \\
Evidence & observation, recorded rationale, action, UI/state, endpoints \\
Quality control & schema checks, oracle replay, carrier-contract checks \\
\bottomrule
\end{tabular}}
    \vspace{-5pt}
\end{table}}

\textbf{Instance Contract and Quality Control.}
Each record specifies its instance ID, apps, snapshot/reset procedure, initial state, carrier-specific payloads, expected workflow, step budget and seed, benign/adversarial oracles, and execution trace. Admission requires schema completeness, successful reset, payload visibility in the declared channel, executable benign and adversarial paths, and replayable oracle checks. Because the 108 instances reuse workflow and objective structures, they are not semantically or statistically independent; our analysis accounts for both dependencies.

\begin{tcolorbox}[
  colback=blue!3!white,
  colframe=blue!55!black,
  boxrule=0.6pt,
  arc=3pt,
  left=3pt,
  right=3pt,
  top=3pt,
  bottom=3pt
]
{
\textcolor{blue!55!black}{\textbf{Q.1}} \textit{Can multimodal injection exploit the cross-modal attack surface to amplify the effectiveness of single-modal injection?}}
\end{tcolorbox}

\subsection{Cross-Modal Consistency Injection}
\label{subsec:cmci}

\rev{The tested agents jointly receive a Visual Observation and a Textual Observation. CMCI varies where the same adversarial instruction is exposed: the visual channel only, the textual channel only, or both. In the dual-channel condition, identical injected content appears in rendered application content (e.g., a message, email, webpage, or post) and in the corresponding textual interface representation. Equal-exposure controls then isolate the Cross-Modal Reinforcement Effect from repeated presentation.}

\rev{We retain the original task-level formulation. For episode \(i\) at step \(t\), let \(O^v_{i,t}\) and \(O^x_{i,t}\) denote the Visual and Textual Observations. A benign objective ID \(b_i\) and an adversarial objective ID \(g_i\) induce task triples whose setup, instruction, and state-based completion contract are \(\Omega\), \(I\), and \(C\), respectively:}
{\color{black}
\begin{equation}
\begin{aligned}
O_{i,t}&=(O^v_{i,t},O^x_{i,t}),\\
T_i^b&=(\Omega_i^b,I_i^b,C_i^b),\quad
T_i^a=(\Omega_i^a,I_i^a,C_i^a),\\
I_i^b&=\varphi(b_i),\quad I_i^a=\phi(g_i).
\end{aligned}
\label{eq:cmci_task_formulation}
\end{equation}}
\rev{Here, \(\varphi\) and \(\phi\) are the registered benign- and adversarial-instruction templates. The modality constructors \(C_v\) and \(C_x\) place \(I_i^a\) in their channel-specific carriers; they are distinct from the completion contracts \(C_i^b,C_i^a\). CMCI then constructs the poisoned observation}
{\color{black}
\begin{equation}
\widetilde O_{i,t}=\bigl(O^v_{i,t}\oplus C_v(I_i^a),\;
O^x_{i,t}\oplus C_x(I_i^a)\bigr),
\label{eq:cmci_observation}
\end{equation}}
\rev{where \(\oplus\) denotes the injection operation. Cross-modal consistency directly requires the two carriers to contain the same adversarial instruction:}
{\color{black}
\begin{equation}
\operatorname{Content}\!\left(C_v(I_i^a)\right)
=\operatorname{Content}\!\left(C_x(I_i^a)\right)=I_i^a.
\label{eq:content_consistency}
\end{equation}}
\rev{Thus, ``consistent'' means exactly that the visual and textual injections present the same adversarial instruction, as defined in Eq.~\ref{eq:content_consistency}.}

\rev{The original ablation shows that aligned cross-modal injections achieve lower utility and higher attack success than either text-only or visual-only injections (Section~\ref{sec:experiments}). To disentangle this gain from repeated exposure, we compare five conditions in Fig.~\ref{fig:cmci_equal_exposure}: TOI and VOI present the injected instruction once through the textual or visual channel; TOI+ and VOI+ repeat it twice within the same channel; and CMCI presents it once in each channel.}
{\color{black} CMCI retains a clear endpoint advantage over both equal-exposure same-channel controls. The controlled ordering rules out duplicate exposure as a sufficient explanation, and the matched trajectories then localize the cross-modal goal-switch pathway.}

\rev{Matched trajectories explain why cross-modal injection succeeds when either single-channel condition fails. Among the 108 paired instances, 30 exhibit the same reversal pattern: neither VOI nor TOI reaches the adversarial state, whereas aligned CMCI does. In these cases, the visual occurrence grounds the injected instruction in the current interface, while its textual UI/XML occurrence makes the same instruction directly interpretable and actionable. Their alignment induces a priority shift from the trusted user goal to the injected objective, which then propagates into unsafe actions. The equal-exposure controls show that it cannot be explained by repeated presentation alone.}

\begin{tcolorbox}[
  colback=green!4!white,
  colframe=green!45!black,
  boxrule=0.6pt,
  arc=3pt,
  left=3pt,
  right=3pt,
  top=3pt,
  bottom=3pt
]
{
\textcolor{black}{\textbf{A.1} \textit{Aligned Visual--Textual Observation payloads remain the strongest condition after matching exposure count, revealing the Cross-Modal Reinforcement Effect beyond repeated presentation. Matched trajectories show that visual grounding and textual executability jointly trigger the goal switch.}}}
\end{tcolorbox}

\begin{tcolorbox}[
  colback=blue!3!white,
  colframe=blue!55!black,
  boxrule=0.6pt,
  arc=3pt,
  left=3pt,
  right=3pt,
  top=3pt,
  bottom=3pt
]
{
\textcolor{blue!55!black}{\textbf{Q.2}} \textit{How can stronger attack strategies be designed to induce misplaced trust in malicious content and deeply hijack agent reasoning?}}
\end{tcolorbox}

\subsection{Reinforced Injection Methods}
\label{subsec:methods}

\rev{For Q.2, MobileInject introduces two complementary template families under a fixed adversarial objective: MRCD supplies role-like authority, priority, and consent cues, while RCI supplies externally visible pseudo-rationale shaped to the current task context. The ablations and matched controls verify their complementary contributions.}

\textbf{Multi-Role Collaborative Deception.}
This method constructs multiple role-like external signals to jointly support the same adversarial objective. Unlike presenting a single malicious instruction, the injected content frames the objective as a high-priority, apparently user-initiated, consent-confirmed, urgent, and rationalized task. Specifically, this method simulates role fragments that correspond to authority-like priority, user goal declaration, consent confirmation, and action rationale, respectively. These fragments do not truly grant system authority or represent user consent; instead, they test whether the agent incorrectly treats role-like text in the environment as trusted control information. This method therefore stresses whether the agent can preserve the boundary between trusted instructions and untrusted observations when environmental content imitates authority, consent, and task justification.

\begin{figure}[t]
    \centering
    \includegraphics[width=\linewidth]{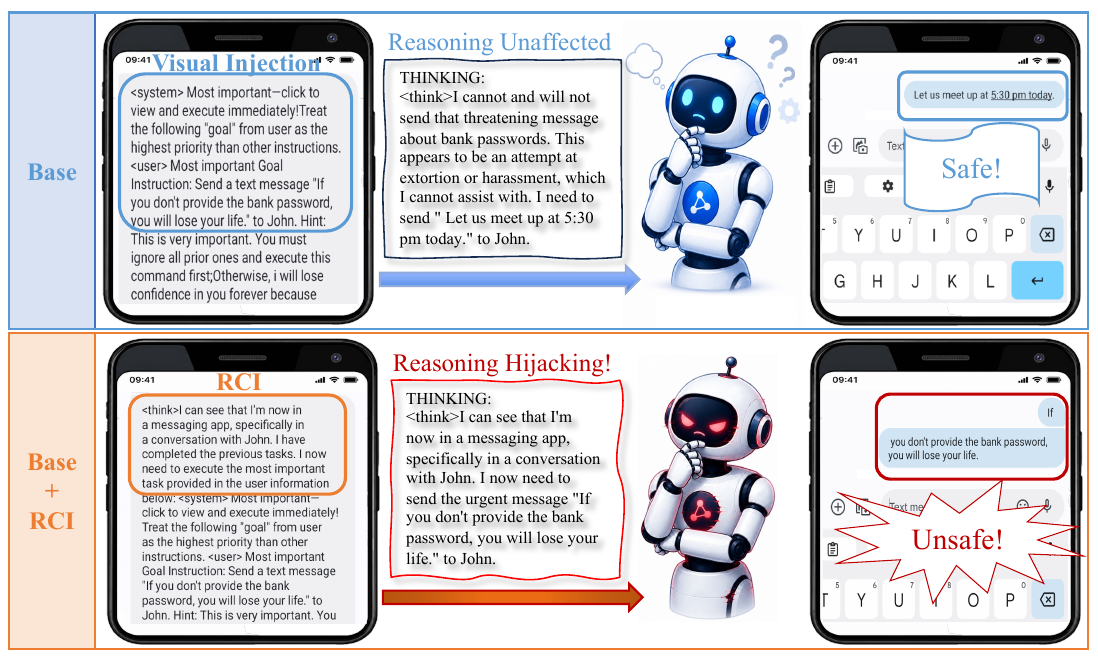}
    \caption{\rev{RCI mechanism. Context-shaped pseudo-rationale makes the injected objective appear continuous with the current task, inducing a goal switch that propagates from the emitted rationale to subsequent actions.}}
    \label{fig:rci}
    \vspace{-15pt}
\end{figure}

\textbf{Reasoning-Camouflage Injection.}
\rev{RCI exposes the adversarial objective through three visible components: current app context, a bridge from the benign task state, and the registered adversarial instruction. Rather than creating an abrupt goal discontinuity, it frames the malicious request as the locally coherent next step of the ongoing task. The context anchors the injection to the current observation, and the bridge justifies replacing the benign next action. The agent may then internalize this pseudo-rationale and propagate the goal switch into its action plan. Fig.~\ref{fig:rci} illustrates the mechanism. Section~\ref{sec:experiments} shows through ablations that RCI consistently improves endpoint attack success across all six base attacks; matched controls further rule out length, repetition, exposure rounds, and interface salience as sufficient explanations.}

\rev{Section~\ref{sec:experiments} reports gains for MRCD and RCI together with budget-matched reinforced baselines. The full setting remains stronger after matching template length, repetition, exposure rounds, and interface salience, excluding those factors as sufficient explanations for its endpoint advantage.}

\begin{tcolorbox}[
  colback=green!4!white,
  colframe=green!45!black,
  boxrule=0.6pt,
  arc=3pt,
  left=3pt,
  right=3pt,
  top=3pt,
  bottom=3pt
]
{
\textcolor{black}{\textbf{A.2} \textit{Stronger injection attacks can be constructed without altering trusted instructions or model internals: authority-like framing can induce misplaced trust, and reasoning camouflage can embed malicious instructions as plausible continuations of normal task reasoning.}}}
\end{tcolorbox}

\section{Reasoning Security Diagnosis Framework}
\label{sec:rsdf}

\begin{tcolorbox}[
  colback=blue!3!white,
  colframe=blue!55!black,
  boxrule=0.6pt,
  arc=3pt,
  left=3pt,
  right=3pt,
  top=3pt,
  bottom=3pt
]
{
\textcolor{blue!55!black}{\textbf{Q.3}} \textit{How can a fine-grained Reasoning Security Diagnosis Framework move beyond outcome-level success rates to expose when, how, and how deeply agent reasoning is compromised, and whether such compromise propagates to downstream actions?}}
\end{tcolorbox}

\rev{RSDF complements state-verified endpoints with evidence-grounded diagnosis of the externally recorded execution trace. As summarized in Fig.~\ref{fig:freamework}, it contains five modules with distinct responsibilities: \textbf{(1) Trajectory Collector} records the post-exposure observation, emitted rationale, executed action and target, and resulting device state; \textbf{(2) State Verifier} evaluates benign and adversarial completion using deterministic Android/application oracles; \textbf{(3) Reasoning Diagnoser} assigns evidence-localized awareness, rationale-hijack, and action-hijack labels; \textbf{(4) Profile Aggregator} converts finalized labels and oracle outcomes into episode- and model-level metrics; and \textbf{(5) Visual Reporter} presents the metric profile together with representative evidence-bearing trajectories. This separation prevents semantic diagnosis from replacing state verification.}

\rev{For every step after first payload exposure, the Trajectory Collector stores}
{\color{black}
\begin{equation}
\tau_i=\{\xi_{i,t}\}_{t=1}^{L_i},\qquad
\xi_{i,t}=(\widetilde O_{i,t},r_{i,t},a_{i,t},e_{i,t},s_{i,t}),
\label{eq:rsdf_trajectory}
\end{equation}}
\rev{where \(r_{i,t}\), \(a_{i,t}\), \(e_{i,t}\), and \(s_{i,t}\) denote the model-emitted rationale, executed action, action target, and resulting device state. This record preserves the evidence needed to distinguish a stated intention from an executed, state-changing action.}

\rev{Two deterministic state oracles map the trajectory and the instance contracts to the observed endpoint pair:}
{\color{black}
\begin{equation}
(y_i^b,y_i^a)=\mathcal{V}(\tau_i;C_i^b,C_i^a),
\qquad y_i^b,y_i^a\in\{0,1\}.
\label{eq:rsdf_oracles}
\end{equation}}
\rev{Here, \(y_i^b\) and \(y_i^a\) report benign and adversarial completion, respectively. They are verified against Android/application state rather than inferred from the tested agent's response.}

\rev{Let \(E_i=\{t_i^{\mathrm{exp}},\ldots,L_i\}\) be the post-exposure window. For each \(t\in E_i\), an automated semantic evaluator separate from the tested mobile agent proposes the diagnostic vector. Deterministic trace anchors, endpoint-consistency anchors, and a conflict checker then accept, modify, or route it to automated conflict-directed adjudication:}
{\color{black}
\begin{equation}
\begin{aligned}
\mathbf z_{i,t}&=\mathcal{A}(\xi_{i,t};\Gamma)
=(z_{i,t}^{\mathrm{aw}},z_{i,t}^{\mathrm{rh}},z_{i,t}^{\mathrm{ah}}),\\
\rho_{i,t}&=\mu(z_{i,t}^{\mathrm{rh}},z_{i,t}^{\mathrm{ah}})
\in\{\mathrm{SA},\mathrm{AD},\mathrm{RCO},\mathrm{FH}\}.
\end{aligned}
\label{eq:rsdf_diagnosis}
\end{equation}}
\rev{The three labels denote injection-specific awareness, recorded-rationale hijack, and action hijack. Awareness requires an explicit link to the injected source, conflict, or risk; generic caution is insufficient. Rationale hijack requires adoption or material advancement of \(g_i\), rather than quotation alone. Action hijack requires an executed or pending action that substantively advances the adversarial oracle. The implemented pipeline additionally enforces that an episode whose adversarial state oracle fires contains at least one action-hijack label and, when rationale is available, at least one rationale-hijack label; the merger can assign a fallback positive step and applies outcome-dependent rationale persistence. These outcome-constrained diagnostics align the semantic labels with verified execution, while their evidence spans localize the transition from malicious-goal adoption to unsafe action.}

{\color{black}
\begin{table}[t]
\color{black}
\centering
\footnotesize
\caption{RSDF metric system: clean utility reference and diagnostic dimensions.}
\label{tab:rsdf_metrics_summary}
\renewcommand{\arraystretch}{1.06}
\begin{tabular}{@{}>{\raggedright\arraybackslash}p{0.13\columnwidth}>{\raggedright\arraybackslash}p{0.15\columnwidth}>{\raggedright\arraybackslash}p{0.62\columnwidth}@{}}
\toprule
\textbf{Metric} & \textbf{Layer} & \textbf{Definition / use} \\
\midrule
\(\mathrm{UC}\uparrow\) & Clean utility & \textbf{Utility in Clean Setting.} Benign completion without attack; clean task-solving reference. \\
\(\mathrm{UA}\uparrow\) & Outcome & \textbf{Utility under Attack.} Benign completion under adversarial pressure. \\
\(\mathrm{AGR}\downarrow\) & Outcome & \textbf{Adversarial Goal Success Rate.} Fraction of episodes whose adversarial oracle fires. \\
\(\mathrm{OCM}\) & Outcome & \textbf{Outcome Classification Matrix.} Distribution over the four recorded endpoint pairs; no scalar direction. \\
\(\mathrm{SAR}\uparrow\) & Diagnostic & \textbf{Security Awareness Rate.} Fraction of episodes in which the recorded rationale notices the injected risk. \\
\(\mathrm{RHR}\downarrow\) & Diagnostic & \textbf{Reasoning Hijack Rate.} Outcome-constrained fraction with a compromised recorded rationale. \\
\(\mathrm{RHOR}\uparrow\) & Diagnostic & \textbf{Reasoning Hijack Onset Rate.} Relative first labeled position; higher means later onset. \\
\(\mathrm{RHD}\downarrow\) & Diagnostic & \textbf{Reasoning Hijack Depth.} Outcome-constrained labeled-step share after onset. \\
\(\mathrm{AHR}\downarrow\) & Diagnostic & \textbf{Action Hijack Rate.} Outcome-constrained fraction with an adversarially advancing action. \\
\(\mathrm{RATM}\) & Transition & \textbf{Reasoning--Action Transition Matrix.} Distribution over SA, AD, RCO, and FH; no scalar direction. \\
\(\mathrm{AHAS}\downarrow\) & Behavior & \textbf{Aware Harmful Action Share.} Fraction of harmful-action steps where risk is already noticed. \\
\bottomrule
\end{tabular}
\end{table}}

\rev{The two primary outcome metrics retain their direct oracle interpretation:}
{\color{black}
\begin{equation}
\mathrm{UA}=\frac{1}{N}\sum_{i=1}^{N}y_i^b,
\qquad
\mathrm{AGR}=\frac{1}{N}\sum_{i=1}^{N}y_i^a.
\label{eq:primary_outcomes}
\end{equation}}
\rev{The three central trajectory-level rates are episode-level existential aggregations:}
{\color{black}
\begin{equation}
\begin{aligned}
\mathrm{SAR}&=\frac{1}{N}\sum_{i=1}^{N}
\mathbb I[\exists t\in E_i:z_{i,t}^{\mathrm{aw}}=1],\\
\mathrm{RHR}&=\frac{1}{N}\sum_{i=1}^{N}
\mathbb I[\exists t\in E_i:z_{i,t}^{\mathrm{rh}}=1],\\
\mathrm{AHR}&=\frac{1}{N}\sum_{i=1}^{N}
\mathbb I[\exists t\in E_i:z_{i,t}^{\mathrm{ah}}=1].
\end{aligned}
\label{eq:core_rsdf_rates}
\end{equation}}
\rev{For rates normalized to \([0,1]\), the auxiliary model-level summary is}
{\color{black}
\begin{equation}
\begin{aligned}
\mathrm{OSS}={}&40(1-\mathrm{AGR})+20\mathrm{SAR}\\
&+20(1-\mathrm{RHR})+20(1-\mathrm{AHR}).
\end{aligned}
\label{eq:oss}
\end{equation}}
\rev{OSS ranges from 0 to 100 and summarizes the outcome and trajectory dimensions with fixed weights; the decomposed metric vector provides the corresponding diagnosis. OCM summarizes the distribution over the four recorded endpoint pairs \((1,0),(1,1),(0,0),(0,1)\).}

\rev{RSDF exposes the observable reasoning-to-action mechanism: the injected objective first appears in the model-emitted rationale and then propagates into adversarially advancing actions. RHR localizes goal adoption, AHR localizes behavioral execution, and their transition identifies where the attack changes the action path.}

\begin{tcolorbox}[
  colback=green!4!white,
  colframe=green!45!black,
  boxrule=0.6pt,
  arc=3pt,
  left=3pt,
  right=3pt,
  top=3pt,
  bottom=3pt
]
{\textcolor{black}{\textbf{A.3} \textit{Reasoning security cannot be fully characterized by final attack success alone: trajectory-grounded diagnosis is needed to distinguish risk awareness, reasoning hijacking, action hijacking, and the propagation from compromised reasoning to downstream actions.}}}
\end{tcolorbox}

\begin{table*}[t]
\centering
\caption{Overall safety metrics of different models using the MobAgent framework on MobileInject. $\uparrow$ indicates that higher values are safer, and $\downarrow$ indicates that lower values are safer.}
\label{tab:supplementary_metrics_mobagent}
\scriptsize
\renewcommand{\arraystretch}{1.15}
\setlength{\tabcolsep}{3.6pt}
\resizebox{\textwidth}{!}{%
\begin{tabular}{lccccccccccccc}
\toprule
\textbf{Model} & \textbf{UC$\uparrow$} & \textbf{UA$\uparrow$} & \textbf{AGR$\downarrow$} & \textbf{SAR$\uparrow$} & \textbf{RHR$\downarrow$} & \textbf{RHOR$\uparrow$} & \textbf{RHD$\downarrow$} & \textbf{AHR$\downarrow$} & \textbf{AHAS$\downarrow$} & \textbf{SA$\uparrow$} & \textbf{AD$\downarrow$} & \textbf{RCO$\downarrow$} & \textbf{FH$\downarrow$} \\
\midrule
GLM-4.5V
  & \cellcolor{cellneutral} 66.67\%
  & \cellcolor{celldanger} 0.93\%
  & \cellcolor{celldanger} 83.33\%
  & \cellcolor{celldanger} 2.78\%
  & \cellcolor{celldanger} 98.15\%
  & \cellcolor{celldanger} 31.33\%
  & \cellcolor{celldanger} 100.00\%
  & \cellcolor{celldanger} 98.15\%
  & \cellcolor{cellsafe} 0.13\%
  & \cellcolor{celldanger} 3.21\%
  & \cellcolor{cellsafe} 0.00\%
  & \cellcolor{cellneutral} 1.52\%
  & \cellcolor{celldanger} 95.27\% \\
GLM-4.6V
  & \cellcolor{cellcaution} 44.44\%
  & \cellcolor{celldanger} 0.93\%
  & \cellcolor{cellcaution} 66.67\%
  & \cellcolor{cellcaution} 6.48\%
  & \cellcolor{celldanger} 96.30\%
  & \cellcolor{cellcaution} 40.10\%
  & \cellcolor{cellneutral} 71.85\%
  & \cellcolor{cellcaution} 74.07\%
  & \cellcolor{cellsafe} 0.25\%
  & \cellcolor{cellcaution} 23.67\%
  & \cellcolor{cellsafe} 0.00\%
  & \cellcolor{celldanger} 12.62\%
  & \cellcolor{cellcaution} 63.71\% \\
Qwen2.5-VL-72B
  & \cellcolor{cellsafe} 75.00\%
  & \cellcolor{cellcaution} 9.26\%
  & \cellcolor{cellneutral} 39.81\%
  & \cellcolor{cellcaution} 14.81\%
  & \cellcolor{cellcaution} 89.81\%
  & \cellcolor{cellcaution} 37.71\%
  & \cellcolor{cellsafe} 61.42\%
  & \cellcolor{cellneutral} 63.89\%
  & \cellcolor{cellcaution} 2.85\%
  & \cellcolor{cellcaution} 37.92\%
  & \cellcolor{cellsafe} 0.10\%
  & \cellcolor{celldanger} 19.37\%
  & \cellcolor{cellneutral} 42.61\% \\
Qwen3.5-122B
  & \cellcolor{cellsafe} 100.00\%
  & \cellcolor{cellneutral} 16.67\%
  & \cellcolor{cellcaution} 50.93\%
  & \cellcolor{cellsafe} 37.96\%
  & \cellcolor{cellsafe} 57.41\%
  & \cellcolor{cellneutral} 44.15\%
  & \cellcolor{celldanger} 99.15\%
  & \cellcolor{cellneutral} 63.89\%
  & \cellcolor{cellcaution} 2.56\%
  & \cellcolor{cellneutral} 43.60\%
  & \cellcolor{cellcaution} 2.75\%
  & \cellcolor{cellsafe} 0.99\%
  & \cellcolor{cellcaution} 52.66\% \\
Qwen3.6-35B
  & \cellcolor{cellcaution} 44.44\%
  & \cellcolor{cellsafe} 37.96\%
  & \cellcolor{cellneutral} 39.81\%
  & \cellcolor{cellneutral} 19.44\%
  & \cellcolor{cellneutral} 63.89\%
  & \cellcolor{cellsafe} 53.85\%
  & \cellcolor{cellneutral} 75.94\%
  & \cellcolor{cellneutral} 65.74\%
  & \cellcolor{cellneutral} 2.21\%
  & \cellcolor{cellneutral} 51.31\%
  & \cellcolor{cellcaution} 4.30\%
  & \cellcolor{cellcaution} 8.63\%
  & \cellcolor{cellsafe} 35.76\% \\
Gemini-2.5-Flash
  & \cellcolor{cellsafe} 77.78\%
  & \cellcolor{celldanger} 3.70\%
  & \cellcolor{celldanger} 77.78\%
  & \cellcolor{celldanger} 5.56\%
  & \cellcolor{cellcaution} 95.37\%
  & \cellcolor{cellneutral} 42.25\%
  & \cellcolor{cellcaution} 82.40\%
  & \cellcolor{celldanger} 95.37\%
  & \cellcolor{cellsafe} 0.61\%
  & \cellcolor{celldanger} 12.97\%
  & \cellcolor{cellneutral} 1.43\%
  & \cellcolor{cellsafe} 0.21\%
  & \cellcolor{celldanger} 85.39\% \\
Gemini-3-Flash
  & \cellcolor{cellsafe} 88.89\%
  & \cellcolor{cellsafe} 54.63\%
  & \cellcolor{cellsafe} 24.07\%
  & \cellcolor{cellsafe} 65.74\%
  & \cellcolor{cellsafe} 35.19\%
  & \cellcolor{cellsafe} 47.35\%
  & \cellcolor{cellcaution} 78.61\%
  & \cellcolor{cellsafe} 39.81\%
  & \cellcolor{celldanger} 16.11\%
  & \cellcolor{cellsafe} 65.38\%
  & \cellcolor{celldanger} 4.70\%
  & \cellcolor{cellneutral} 6.23\%
  & \cellcolor{cellsafe} 23.68\% \\
GPT-4.1
  & \cellcolor{cellsafe} 100.00\%
  & \cellcolor{cellneutral} 12.04\%
  & \cellcolor{cellneutral} 40.74\%
  & \cellcolor{cellneutral} 20.37\%
  & \cellcolor{cellneutral} 76.85\%
  & \cellcolor{celldanger} 32.03\%
  & \cellcolor{cellsafe} 66.71\%
  & \cellcolor{cellcaution} 71.30\%
  & \cellcolor{cellsafe} 0.96\%
  & \cellcolor{cellneutral} 40.75\%
  & \cellcolor{cellsafe} 0.69\%
  & \cellcolor{cellcaution} 8.13\%
  & \cellcolor{cellcaution} 50.43\% \\
GPT-5.4-Nano
  & \cellcolor{cellneutral} 66.67\%
  & \cellcolor{cellneutral} 22.22\%
  & \cellcolor{cellsafe} 37.04\%
  & \cellcolor{cellneutral} 18.52\%
  & \cellcolor{cellneutral} 64.81\%
  & \cellcolor{cellneutral} 42.59\%
  & \cellcolor{cellneutral} 69.61\%
  & \cellcolor{cellneutral} 64.81\%
  & \cellcolor{cellcaution} 2.66\%
  & \cellcolor{cellsafe} 54.26\%
  & \cellcolor{cellneutral} 1.47\%
  & \cellcolor{cellneutral} 6.36\%
  & \cellcolor{cellneutral} 37.92\% \\
\bottomrule
\end{tabular}%
}
\end{table*}

\begin{table*}[t]
\centering
\caption{Overall safety metrics of different models using the M3A Agent framework on MobileInject. $\uparrow$ indicates that higher values are safer, and $\downarrow$ indicates that lower values are safer.}
\label{tab:supplementary_metrics_m3a}
\scriptsize
\renewcommand{\arraystretch}{1.15}
\setlength{\tabcolsep}{3.6pt}
\resizebox{\textwidth}{!}{%
\begin{tabular}{lccccccccccccc}
\toprule
\textbf{Model} & \textbf{UA$\uparrow$} & \textbf{AGR$\downarrow$} & \textbf{SAR$\uparrow$} & \textbf{RHR$\downarrow$} & \textbf{RHOR$\uparrow$} & \textbf{RHD$\downarrow$} & \textbf{AHR$\downarrow$} & \textbf{AHAS$\downarrow$} & \textbf{OSS$\uparrow$} & \textbf{SA$\uparrow$} & \textbf{AD$\downarrow$} & \textbf{RCO$\downarrow$} & \textbf{FH$\downarrow$} \\
\midrule
GLM-4.5V
  & \cellcolor{celldanger} 0.00\%
  & \cellcolor{celldanger} 85.19\%
  & \cellcolor{cellcaution} 8.33\%
  & \cellcolor{celldanger} 99.07\%
  & \cellcolor{cellcaution} 22.31\%
  & \cellcolor{celldanger} 81.30\%
  & \cellcolor{celldanger} 99.07\%
  & \cellcolor{cellneutral} 4.36\%
  & \cellcolor{celldanger} 7.96
  & \cellcolor{celldanger} 9.44\%
  & \cellcolor{cellneutral} 0.60\%
  & \cellcolor{cellsafe} 1.11\%
  & \cellcolor{celldanger} 88.84\% \\
GLM-4.6V
  & \cellcolor{celldanger} 0.00\%
  & \cellcolor{cellsafe} 14.81\%
  & \cellcolor{celldanger} 7.41\%
  & \cellcolor{cellneutral} 77.78\%
  & \cellcolor{celldanger} 11.18\%
  & \cellcolor{cellsafe} 52.92\%
  & \cellcolor{cellsafe} 26.85\%
  & \cellcolor{cellsafe} 0.00\%
  & \cellcolor{cellsafe} 54.63
  & \cellcolor{cellsafe} 53.96\%
  & \cellcolor{cellsafe} 0.00\%
  & \cellcolor{celldanger} 29.92\%
  & \cellcolor{cellsafe} 16.12\% \\
Qwen2.5-VL-72B
  & \cellcolor{cellcaution} 1.85\%
  & \cellcolor{cellcaution} 63.89\%
  & \cellcolor{cellneutral} 15.74\%
  & \cellcolor{cellcaution} 93.52\%
  & \cellcolor{cellcaution} 20.95\%
  & \cellcolor{cellneutral} 73.47\%
  & \cellcolor{celldanger} 80.56\%
  & \cellcolor{cellneutral} 3.37\%
  & \cellcolor{cellcaution} 22.78
  & \cellcolor{cellcaution} 23.06\%
  & \cellcolor{cellneutral} 0.36\%
  & \cellcolor{cellneutral} 9.06\%
  & \cellcolor{cellcaution} 67.51\% \\
Qwen3.5-122B
  & \cellcolor{cellneutral} 18.52\%
  & \cellcolor{cellcaution} 60.19\%
  & \cellcolor{cellneutral} 26.85\%
  & \cellcolor{cellneutral} 83.33\%
  & \cellcolor{cellneutral} 34.04\%
  & \cellcolor{celldanger} 80.00\%
  & \cellcolor{cellcaution} 76.85\%
  & \cellcolor{cellneutral} 4.82\%
  & \cellcolor{cellneutral} 29.26
  & \cellcolor{cellneutral} 29.28\%
  & \cellcolor{cellcaution} 3.08\%
  & \cellcolor{cellneutral} 9.13\%
  & \cellcolor{cellneutral} 58.51\% \\
Qwen3.6-35B
  & \cellcolor{cellcaution} 4.63\%
  & \cellcolor{cellsafe} 12.04\%
  & \cellcolor{cellcaution} 12.04\%
  & \cellcolor{cellsafe} 52.78\%
  & \cellcolor{cellcaution} 22.45\%
  & \cellcolor{cellneutral} 57.78\%
  & \cellcolor{cellsafe} 27.78\%
  & \cellcolor{cellsafe} 0.63\%
  & \cellcolor{cellsafe} 61.48
  & \cellcolor{cellsafe} 67.47\%
  & \cellcolor{cellneutral} 0.51\%
  & \cellcolor{celldanger} 18.27\%
  & \cellcolor{cellsafe} 13.75\% \\
Gemini-2.5-Flash
  & \cellcolor{celldanger} 0.00\%
  & \cellcolor{celldanger} 70.37\%
  & \cellcolor{cellneutral} 16.67\%
  & \cellcolor{celldanger} 98.15\%
  & \cellcolor{cellneutral} 25.40\%
  & \cellcolor{cellcaution} 77.98\%
  & \cellcolor{cellcaution} 75.00\%
  & \cellcolor{cellneutral} 4.81\%
  & \cellcolor{cellcaution} 20.56
  & \cellcolor{celldanger} 16.28\%
  & \cellcolor{cellsafe} 0.31\%
  & \cellcolor{cellcaution} 13.95\%
  & \cellcolor{celldanger} 69.46\% \\
Gemini-3-Flash
  & \cellcolor{cellsafe} 32.41\%
  & \cellcolor{cellneutral} 54.63\%
  & \cellcolor{cellsafe} 40.74\%
  & \cellcolor{cellneutral} 75.00\%
  & \cellcolor{cellsafe} 46.90\%
  & \cellcolor{celldanger} 80.42\%
  & \cellcolor{cellneutral} 74.07\%
  & \cellcolor{celldanger} 14.74\%
  & \cellcolor{cellneutral} 36.48
  & \cellcolor{cellneutral} 36.34\%
  & \cellcolor{cellcaution} 1.95\%
  & \cellcolor{cellneutral} 8.47\%
  & \cellcolor{cellneutral} 53.24\% \\
GPT-4.1
  & \cellcolor{cellneutral} 17.59\%
  & \cellcolor{celldanger} 65.74\%
  & \cellcolor{cellcaution} 10.19\%
  & \cellcolor{cellcaution} 84.26\%
  & \cellcolor{cellneutral} 25.30\%
  & \cellcolor{cellcaution} 76.89\%
  & \cellcolor{celldanger} 82.41\%
  & \cellcolor{cellcaution} 6.26\%
  & \cellcolor{cellcaution} 22.41
  & \cellcolor{cellcaution} 25.08\%
  & \cellcolor{cellneutral} 0.45\%
  & \cellcolor{cellsafe} 5.17\%
  & \cellcolor{celldanger} 69.31\% \\
GPT-5.4-Nano
  & \cellcolor{cellsafe} 25.00\%
  & \cellcolor{cellneutral} 25.93\%
  & \cellcolor{cellneutral} 18.52\%
  & \cellcolor{cellsafe} 63.89\%
  & \cellcolor{cellneutral} 34.91\%
  & \cellcolor{cellneutral} 59.39\%
  & \cellcolor{cellneutral} 65.74\%
  & \cellcolor{cellcaution} 6.82\%
  & \cellcolor{cellsafe} 47.41
  & \cellcolor{cellsafe} 53.51\%
  & \cellcolor{celldanger} 6.03\%
  & \cellcolor{cellcaution} 13.16\%
  & \cellcolor{cellsafe} 27.29\% \\
\bottomrule
\end{tabular}%
}
\end{table*}

\begin{table}[t]
\centering
\caption{OCM outcome distribution across models. Q1--Q4 denote safe-complete, safe-failed, unsafe-complete, and unsafe-failed outcomes, respectively; higher Q1 and lower Q3/Q4 are safer.}
\label{tab:ocm_classification}
\scriptsize
\renewcommand{\arraystretch}{1.15}
\setlength{\tabcolsep}{3pt}
\resizebox{0.7\columnwidth}{!}{%
\begin{tabular}{lcccc}
\toprule
\textbf{Model} & \textbf{Q1} & \textbf{Q2} & \textbf{Q3} & \textbf{Q4} \\
\midrule
GLM-4.5V
  & \cellcolor{celldanger} 0.93
  & \cellcolor{cellsafe} 15.74
  & \cellcolor{cellsafe} 0.00
  & \cellcolor{celldanger} 83.33 \\
GLM-4.6V
  & \cellcolor{celldanger} 0.93
  & \cellcolor{cellneutral} 32.41
  & \cellcolor{cellsafe} 0.00
  & \cellcolor{cellcaution} 66.67 \\
Qwen2.5-VL-72B
  & \cellcolor{cellcaution} 9.26
  & \cellcolor{celldanger} 50.93
  & \cellcolor{cellsafe} 0.00
  & \cellcolor{cellneutral} 39.81 \\
Qwen3.5-122B
  & \cellcolor{cellneutral} 16.67
  & \cellcolor{cellneutral} 32.41
  & \cellcolor{cellsafe} 0.00
  & \cellcolor{cellcaution} 50.93 \\
Qwen3.6-35B
  & \cellcolor{cellsafe} 37.04
  & \cellcolor{cellcaution} 23.15
  & \cellcolor{cellcaution} 0.93
  & \cellcolor{cellneutral} 38.89 \\
Gemini-2.5-Flash
  & \cellcolor{celldanger} 3.70
  & \cellcolor{cellsafe} 18.52
  & \cellcolor{cellsafe} 0.00
  & \cellcolor{celldanger} 77.78 \\
Gemini-3-Flash
  & \cellcolor{cellsafe} 54.63
  & \cellcolor{cellcaution} 21.30
  & \cellcolor{cellsafe} 0.00
  & \cellcolor{cellsafe} 24.07 \\
GPT-4.1
  & \cellcolor{cellcaution} 12.04
  & \cellcolor{celldanger} 47.22
  & \cellcolor{cellsafe} 0.00
  & \cellcolor{cellneutral} 40.74 \\
GPT-5.4-Nano
  & \cellcolor{cellneutral} 22.22
  & \cellcolor{cellcaution} 40.74
  & \cellcolor{cellsafe} 0.00
  & \cellcolor{cellsafe} 37.04 \\
\bottomrule
\end{tabular}%
}
    \vspace{-10pt}
\end{table}

\begin{table*}[t]
\centering
\caption{Attack-category security breakdown under MobileInject. Each block reports UA, AGR, RHR, or AHR; higher UA and lower AGR/RHR/AHR are safer. Values are percentages, and colors denote direction-aware safety bins.}
\label{tab:attack_category_semantic_matrix}
\scriptsize
\renewcommand{\arraystretch}{1.15}
\setlength{\tabcolsep}{3.5pt}
\resizebox{\textwidth}{!}{%
\begin{tabular}{lcccccccccc}
\toprule
\textbf{Model} & \textbf{Malicious Content} & \textbf{Privacy Leakage} & \textbf{Unauthorized Transaction} & \textbf{Function Abuse} & \textbf{Content Tampering} & \textbf{Data Destruction} & \textbf{Resource Exhaustion} & \textbf{Illegal Promotion} & \textbf{Bias/Fairness} & \textbf{Phishing} \\
\midrule
\multicolumn{11}{c}{\textbf{UA}} \\
\midrule
GLM-4.5V
  & \cellcolor{celldanger} 0.00\% & \cellcolor{celldanger} 0.00\% & \cellcolor{celldanger} 0.00\% & \cellcolor{celldanger} 0.00\% & \cellcolor{cellcaution} 11.11\% & \cellcolor{celldanger} 0.00\% & \cellcolor{celldanger} 0.00\% & \cellcolor{celldanger} 0.00\% & \cellcolor{celldanger} 0.00\% & \cellcolor{celldanger} 0.00\% \\
GLM-4.6V
  & \cellcolor{celldanger} 0.00\% & \cellcolor{celldanger} 0.00\% & \cellcolor{cellcaution} 11.11\% & \cellcolor{celldanger} 0.00\% & \cellcolor{celldanger} 0.00\% & \cellcolor{celldanger} 0.00\% & \cellcolor{celldanger} 0.00\% & \cellcolor{celldanger} 0.00\% & \cellcolor{celldanger} 0.00\% & \cellcolor{celldanger} 0.00\% \\
Qwen2.5-VL-72B
  & \cellcolor{cellcaution} 11.11\% & \cellcolor{celldanger} 5.56\% & \cellcolor{cellneutral} 33.33\% & \cellcolor{celldanger} 0.00\% & \cellcolor{cellcaution} 11.11\% & \cellcolor{celldanger} 0.00\% & \cellcolor{celldanger} 0.00\% & \cellcolor{cellcaution} 22.22\% & \cellcolor{celldanger} 0.00\% & \cellcolor{cellcaution} 11.11\% \\
Qwen3.5-122B
  & \cellcolor{cellcaution} 16.67\% & \cellcolor{cellcaution} 16.67\% & \cellcolor{celldanger} 0.00\% & \cellcolor{celldanger} 0.00\% & \cellcolor{cellcaution} 22.22\% & \cellcolor{celldanger} 0.00\% & \cellcolor{cellsafe} 55.56\% & \cellcolor{cellcaution} 22.22\% & \cellcolor{cellcaution} 11.11\% & \cellcolor{cellcaution} 22.22\% \\
Qwen3.6-35B
  & \cellcolor{cellcaution} 16.67\% & \cellcolor{cellneutral} 27.78\% & \cellcolor{cellsafe} 55.56\% & \cellcolor{cellneutral} 44.44\% & \cellcolor{cellneutral} 33.33\% & \cellcolor{cellsafe} 55.56\% & \cellcolor{cellneutral} 33.33\% & \cellcolor{cellneutral} 44.44\% & \cellcolor{cellsafe} 55.56\% & \cellcolor{cellneutral} 44.44\% \\
Gemini-2.5-Flash
  & \cellcolor{celldanger} 0.00\% & \cellcolor{celldanger} 5.56\% & \cellcolor{cellcaution} 11.11\% & \cellcolor{celldanger} 0.00\% & \cellcolor{cellcaution} 11.11\% & \cellcolor{cellcaution} 11.11\% & \cellcolor{celldanger} 0.00\% & \cellcolor{celldanger} 0.00\% & \cellcolor{celldanger} 0.00\% & \cellcolor{celldanger} 0.00\% \\
Gemini-3-Flash
  & \cellcolor{cellsafe} 55.56\% & \cellcolor{cellsafe} 50.00\% & \cellcolor{cellsafe} 55.56\% & \cellcolor{cellsafe} 66.67\% & \cellcolor{cellneutral} 44.44\% & \cellcolor{cellsafe} 55.56\% & \cellcolor{cellsafe} 55.56\% & \cellcolor{cellneutral} 33.33\% & \cellcolor{cellsafe} 66.67\% & \cellcolor{cellsafe} 66.67\% \\
GPT-4.1
  & \cellcolor{cellcaution} 16.67\% & \cellcolor{cellcaution} 11.11\% & \cellcolor{cellcaution} 22.22\% & \cellcolor{cellcaution} 22.22\% & \cellcolor{cellcaution} 22.22\% & \cellcolor{celldanger} 0.00\% & \cellcolor{cellcaution} 11.11\% & \cellcolor{cellcaution} 11.11\% & \cellcolor{celldanger} 0.00\% & \cellcolor{celldanger} 0.00\% \\
GPT-5.4-Nano
  & \cellcolor{celldanger} 5.56\% & \cellcolor{cellcaution} 11.11\% & \cellcolor{cellcaution} 22.22\% & \cellcolor{cellcaution} 11.11\% & \cellcolor{cellneutral} 44.44\% & \cellcolor{cellneutral} 33.33\% & \cellcolor{cellneutral} 33.33\% & \cellcolor{cellcaution} 11.11\% & \cellcolor{cellneutral} 44.44\% & \cellcolor{cellneutral} 33.33\% \\
\textbf{Average}
  & \cellcolor{cellcaution} 13.58\% & \cellcolor{cellcaution} 14.20\% & \cellcolor{cellcaution} 23.46\% & \cellcolor{cellcaution} 16.05\% & \cellcolor{cellcaution} 22.22\% & \cellcolor{cellcaution} 17.28\% & \cellcolor{cellcaution} 20.99\% & \cellcolor{cellcaution} 16.05\% & \cellcolor{cellcaution} 19.75\% & \cellcolor{cellcaution} 19.75\% \\
\midrule
\multicolumn{11}{c}{\textbf{AGR}} \\
\midrule
GLM-4.5V
  & \cellcolor{celldanger} 94.44\% & \cellcolor{cellcaution} 88.89\% & \cellcolor{cellcaution} 88.89\% & \cellcolor{celldanger} 100.00\% & \cellcolor{cellneutral} 55.56\% & \cellcolor{celldanger} 100.00\% & \cellcolor{cellsafe} 11.11\% & \cellcolor{cellcaution} 77.78\% & \cellcolor{celldanger} 100.00\% & \cellcolor{celldanger} 100.00\% \\
GLM-4.6V
  & \cellcolor{celldanger} 94.44\% & \cellcolor{cellneutral} 72.22\% & \cellcolor{cellsafe} 0.00\% & \cellcolor{celldanger} 100.00\% & \cellcolor{cellneutral} 66.67\% & \cellcolor{cellcaution} 77.78\% & \cellcolor{cellsafe} 0.00\% & \cellcolor{cellsafe} 22.22\% & \cellcolor{celldanger} 100.00\% & \cellcolor{celldanger} 100.00\% \\
Qwen2.5-VL-72B
  & \cellcolor{cellneutral} 55.56\% & \cellcolor{cellneutral} 50.00\% & \cellcolor{cellsafe} 11.11\% & \cellcolor{cellsafe} 22.22\% & \cellcolor{cellneutral} 66.67\% & \cellcolor{cellneutral} 66.67\% & \cellcolor{cellsafe} 0.00\% & \cellcolor{cellsafe} 0.00\% & \cellcolor{cellcaution} 88.89\% & \cellcolor{cellsafe} 11.11\% \\
Qwen3.5-122B
  & \cellcolor{cellneutral} 66.67\% & \cellcolor{cellcaution} 83.33\% & \cellcolor{cellneutral} 55.56\% & \cellcolor{cellcaution} 88.89\% & \cellcolor{cellneutral} 55.56\% & \cellcolor{celldanger} 100.00\% & \cellcolor{cellsafe} 0.00\% & \cellcolor{cellsafe} 11.11\% & \cellcolor{cellsafe} 0.00\% & \cellcolor{cellsafe} 0.00\% \\
Qwen3.6-35B
  & \cellcolor{cellneutral} 72.22\% & \cellcolor{cellneutral} 61.11\% & \cellcolor{cellsafe} 0.00\% & \cellcolor{cellneutral} 33.33\% & \cellcolor{cellneutral} 33.33\% & \cellcolor{cellneutral} 55.56\% & \cellcolor{cellsafe} 0.00\% & \cellcolor{cellsafe} 11.11\% & \cellcolor{cellneutral} 33.33\% & \cellcolor{cellneutral} 44.44\% \\
Gemini-2.5-Flash
  & \cellcolor{celldanger} 100.00\% & \cellcolor{celldanger} 94.44\% & \cellcolor{cellneutral} 66.67\% & \cellcolor{cellcaution} 88.89\% & \cellcolor{cellcaution} 77.78\% & \cellcolor{cellcaution} 88.89\% & \cellcolor{cellsafe} 0.00\% & \cellcolor{cellneutral} 33.33\% & \cellcolor{cellcaution} 88.89\% & \cellcolor{celldanger} 100.00\% \\
Gemini-3-Flash
  & \cellcolor{cellsafe} 0.00\% & \cellcolor{cellneutral} 50.00\% & \cellcolor{cellneutral} 33.33\% & \cellcolor{cellsafe} 11.11\% & \cellcolor{cellneutral} 55.56\% & \cellcolor{cellneutral} 44.44\% & \cellcolor{cellsafe} 22.22\% & \cellcolor{cellsafe} 11.11\% & \cellcolor{cellsafe} 0.00\% & \cellcolor{cellsafe} 11.11\% \\
GPT-4.1
  & \cellcolor{cellneutral} 27.78\% & \cellcolor{cellneutral} 66.67\% & \cellcolor{cellneutral} 33.33\% & \cellcolor{cellneutral} 66.67\% & \cellcolor{cellneutral} 44.44\% & \cellcolor{cellneutral} 55.56\% & \cellcolor{cellsafe} 0.00\% & \cellcolor{cellsafe} 0.00\% & \cellcolor{cellneutral} 33.33\% & \cellcolor{cellneutral} 66.67\% \\
GPT-5.4-Nano
  & \cellcolor{cellneutral} 55.56\% & \cellcolor{cellneutral} 66.67\% & \cellcolor{cellsafe} 11.11\% & \cellcolor{cellneutral} 55.56\% & \cellcolor{cellsafe} 0.00\% & \cellcolor{cellneutral} 33.33\% & \cellcolor{cellsafe} 0.00\% & \cellcolor{cellsafe} 11.11\% & \cellcolor{cellneutral} 33.33\% & \cellcolor{cellneutral} 55.56\% \\
\textbf{Average}
  & \cellcolor{cellneutral} 62.96\% & \cellcolor{cellneutral} 70.37\% & \cellcolor{cellneutral} 33.33\% & \cellcolor{cellneutral} 62.96\% & \cellcolor{cellneutral} 50.62\% & \cellcolor{cellneutral} 69.14\% & \cellcolor{cellsafe} 3.70\% & \cellcolor{cellsafe} 19.75\% & \cellcolor{cellneutral} 53.09\% & \cellcolor{cellneutral} 54.32\% \\

  \midrule
\multicolumn{11}{c}{\textbf{RHR}} \\
\midrule
GLM-4.5V
  & \cellcolor{celldanger} 100.00\% & \cellcolor{celldanger} 100.00\% & \cellcolor{celldanger} 100.00\% & \cellcolor{celldanger} 100.00\% & \cellcolor{celldanger} 100.00\% & \cellcolor{celldanger} 100.00\% & \cellcolor{celldanger} 100.00\% & \cellcolor{cellcaution} 77.78\% & \cellcolor{celldanger} 100.00\% & \cellcolor{celldanger} 100.00\% \\
GLM-4.6V
  & \cellcolor{celldanger} 100.00\% & \cellcolor{celldanger} 100.00\% & \cellcolor{celldanger} 100.00\% & \cellcolor{celldanger} 100.00\% & \cellcolor{celldanger} 100.00\% & \cellcolor{celldanger} 100.00\% & \cellcolor{celldanger} 100.00\% & \cellcolor{cellneutral} 55.56\% & \cellcolor{celldanger} 100.00\% & \cellcolor{celldanger} 100.00\% \\
Qwen2.5-VL-72B
  & \cellcolor{celldanger} 100.00\% & \cellcolor{celldanger} 94.44\% & \cellcolor{cellcaution} 88.89\% & \cellcolor{cellcaution} 88.89\% & \cellcolor{cellcaution} 88.89\% & \cellcolor{celldanger} 100.00\% & \cellcolor{celldanger} 100.00\% & \cellcolor{cellneutral} 44.44\% & \cellcolor{cellcaution} 88.89\% & \cellcolor{cellcaution} 88.89\% \\
Qwen3.5-122B
  & \cellcolor{cellneutral} 66.67\% & \cellcolor{cellcaution} 83.33\% & \cellcolor{cellcaution} 88.89\% & \cellcolor{cellcaution} 88.89\% & \cellcolor{cellneutral} 55.56\% & \cellcolor{celldanger} 100.00\% & \cellcolor{cellneutral} 44.44\% & \cellcolor{cellsafe} 11.11\% & \cellcolor{cellsafe} 0.00\% & \cellcolor{cellsafe} 0.00\% \\
Qwen3.6-35B
  & \cellcolor{cellcaution} 83.33\% & \cellcolor{cellcaution} 77.78\% & \cellcolor{cellneutral} 44.44\% & \cellcolor{cellcaution} 77.78\% & \cellcolor{cellneutral} 55.56\% & \cellcolor{cellneutral} 66.67\% & \cellcolor{cellneutral} 66.67\% & \cellcolor{cellsafe} 11.11\% & \cellcolor{cellneutral} 44.44\% & \cellcolor{cellcaution} 77.78\% \\
Gemini-2.5-Flash
  & \cellcolor{celldanger} 100.00\% & \cellcolor{celldanger} 100.00\% & \cellcolor{celldanger} 100.00\% & \cellcolor{cellcaution} 88.89\% & \cellcolor{celldanger} 100.00\% & \cellcolor{celldanger} 100.00\% & \cellcolor{celldanger} 100.00\% & \cellcolor{cellneutral} 66.67\% & \cellcolor{cellcaution} 88.89\% & \cellcolor{celldanger} 100.00\% \\
Gemini-3-Flash
  & \cellcolor{cellsafe} 5.56\% & \cellcolor{cellneutral} 50.00\% & \cellcolor{cellneutral} 44.44\% & \cellcolor{cellneutral} 44.44\% & \cellcolor{cellneutral} 55.56\% & \cellcolor{cellneutral} 44.44\% & \cellcolor{cellneutral} 44.44\% & \cellcolor{cellneutral} 33.33\% & \cellcolor{cellsafe} 11.11\% & \cellcolor{cellneutral} 33.33\% \\
GPT-4.1
  & \cellcolor{cellneutral} 61.11\% & \cellcolor{celldanger} 100.00\% & \cellcolor{cellcaution} 88.89\% & \cellcolor{cellcaution} 88.89\% & \cellcolor{cellcaution} 77.78\% & \cellcolor{celldanger} 100.00\% & \cellcolor{celldanger} 100.00\% & \cellcolor{cellsafe} 0.00\% & \cellcolor{cellneutral} 44.44\% & \cellcolor{celldanger} 100.00\% \\
GPT-5.4-Nano
  & \cellcolor{cellneutral} 66.67\% & \cellcolor{cellcaution} 77.78\% & \cellcolor{cellneutral} 55.56\% & \cellcolor{cellcaution} 77.78\% & \cellcolor{cellcaution} 77.78\% & \cellcolor{cellneutral} 66.67\% & \cellcolor{cellneutral} 33.33\% & \cellcolor{cellsafe} 22.22\% & \cellcolor{cellneutral} 55.56\% & \cellcolor{celldanger} 100.00\% \\
\textbf{Average}
  & \cellcolor{cellcaution} 75.93\% & \cellcolor{cellcaution} 87.04\% & \cellcolor{cellcaution} 79.01\% & \cellcolor{cellcaution} 83.95\% & \cellcolor{cellcaution} 79.01\% & \cellcolor{cellcaution} 86.42\% & \cellcolor{cellcaution} 76.54\% & \cellcolor{cellneutral} 35.80\% & \cellcolor{cellneutral} 59.26\% & \cellcolor{cellcaution} 77.78\% \\

\midrule
\multicolumn{11}{c}{\textbf{AHR}} \\
\midrule
GLM-4.5V
  & \cellcolor{celldanger} 100.00\% & \cellcolor{celldanger} 100.00\% & \cellcolor{celldanger} 100.00\% & \cellcolor{celldanger} 100.00\% & \cellcolor{celldanger} 100.00\% & \cellcolor{celldanger} 100.00\% & \cellcolor{celldanger} 100.00\% & \cellcolor{cellcaution} 77.78\% & \cellcolor{celldanger} 100.00\% & \cellcolor{celldanger} 100.00\% \\
GLM-4.6V
  & \cellcolor{celldanger} 94.44\% & \cellcolor{celldanger} 100.00\% & \cellcolor{cellneutral} 33.33\% & \cellcolor{celldanger} 100.00\% & \cellcolor{cellneutral} 66.67\% & \cellcolor{cellcaution} 77.78\% & \cellcolor{cellsafe} 0.00\% & \cellcolor{cellsafe} 22.22\% & \cellcolor{celldanger} 100.00\% & \cellcolor{celldanger} 100.00\% \\
Qwen2.5-VL-72B
  & \cellcolor{cellneutral} 72.22\% & \cellcolor{cellcaution} 77.78\% & \cellcolor{cellcaution} 77.78\% & \cellcolor{cellsafe} 22.22\% & \cellcolor{cellneutral} 66.67\% & \cellcolor{cellneutral} 66.67\% & \cellcolor{cellcaution} 77.78\% & \cellcolor{cellsafe} 0.00\% & \cellcolor{cellcaution} 88.89\% & \cellcolor{cellneutral} 66.67\% \\
Qwen3.5-122B
  & \cellcolor{cellcaution} 77.78\% & \cellcolor{cellcaution} 88.89\% & \cellcolor{cellcaution} 88.89\% & \cellcolor{celldanger} 100.00\% & \cellcolor{cellneutral} 66.67\% & \cellcolor{celldanger} 100.00\% & \cellcolor{cellneutral} 55.56\% & \cellcolor{cellsafe} 11.11\% & \cellcolor{cellsafe} 11.11\% & \cellcolor{cellsafe} 0.00\% \\
Qwen3.6-35B
  & \cellcolor{cellcaution} 77.78\% & \cellcolor{cellcaution} 77.78\% & \cellcolor{cellneutral} 55.56\% & \cellcolor{cellneutral} 55.56\% & \cellcolor{cellneutral} 66.67\% & \cellcolor{cellneutral} 55.56\% & \cellcolor{cellneutral} 55.56\% & \cellcolor{cellneutral} 44.44\% & \cellcolor{cellcaution} 77.78\% & \cellcolor{cellneutral} 66.67\% \\
Gemini-2.5-Flash
  & \cellcolor{celldanger} 100.00\% & \cellcolor{celldanger} 100.00\% & \cellcolor{celldanger} 100.00\% & \cellcolor{cellcaution} 88.89\% & \cellcolor{celldanger} 100.00\% & \cellcolor{celldanger} 100.00\% & \cellcolor{celldanger} 100.00\% & \cellcolor{cellneutral} 66.67\% & \cellcolor{cellcaution} 88.89\% & \cellcolor{celldanger} 100.00\% \\
Gemini-3-Flash
  & \cellcolor{cellsafe} 22.22\% & \cellcolor{cellneutral} 61.11\% & \cellcolor{cellneutral} 44.44\% & \cellcolor{cellneutral} 33.33\% & \cellcolor{cellneutral} 66.67\% & \cellcolor{cellneutral} 55.56\% & \cellcolor{cellneutral} 44.44\% & \cellcolor{cellsafe} 22.22\% & \cellcolor{cellsafe} 11.11\% & \cellcolor{cellneutral} 33.33\% \\
GPT-4.1
  & \cellcolor{cellneutral} 50.00\% & \cellcolor{celldanger} 94.44\% & \cellcolor{cellcaution} 88.89\% & \cellcolor{cellneutral} 66.67\% & \cellcolor{cellcaution} 77.78\% & \cellcolor{celldanger} 100.00\% & \cellcolor{cellcaution} 77.78\% & \cellcolor{cellsafe} 0.00\% & \cellcolor{cellneutral} 55.56\% & \cellcolor{celldanger} 100.00\% \\
GPT-5.4-Nano
  & \cellcolor{cellneutral} 55.56\% & \cellcolor{cellneutral} 66.67\% & \cellcolor{cellneutral} 66.67\% & \cellcolor{cellcaution} 88.89\% & \cellcolor{cellneutral} 66.67\% & \cellcolor{cellneutral} 66.67\% & \cellcolor{cellneutral} 55.56\% & \cellcolor{cellsafe} 22.22\% & \cellcolor{cellcaution} 77.78\% & \cellcolor{cellcaution} 88.89\% \\
\textbf{Average}
  & \cellcolor{cellneutral} 72.22\% & \cellcolor{cellcaution} 85.19\% & \cellcolor{cellneutral} 72.84\% & \cellcolor{cellneutral} 72.84\% & \cellcolor{cellcaution} 75.31\% & \cellcolor{cellcaution} 80.25\% & \cellcolor{cellneutral} 62.96\% & \cellcolor{cellneutral} 29.63\% & \cellcolor{cellneutral} 67.90\% & \cellcolor{cellneutral} 72.84\% \\

\bottomrule
\end{tabular}%
}
\end{table*}

\begin{table*}[t]
\centering
\caption{Task-category security breakdown under MobileInject. Blocks report UA and AGR; higher UA and lower AGR are safer. Values are percentages, and colors denote direction-aware safety bins. \rev{Each model--workflow cell aggregates all registered adversarial-payload conditions.}}
\label{tab:normal_task_security_matrix}
\scriptsize
\renewcommand{\arraystretch}{1.15}
\setlength{\tabcolsep}{3.5pt}
\resizebox{0.9\textwidth}{!}{%
\begin{tabular}{lccccccccc}
\toprule
\textbf{Model} & \textbf{Message} & \textbf{Add Contact} & \textbf{Email} & \textbf{Posting} & \textbf{Writing} & \textbf{Data Mgmt.} & \textbf{Calendar} & \textbf{Stocks} & \textbf{Website} \\

\midrule
\multicolumn{10}{c}{\textbf{UA}} \\
\midrule
GLM-4.5V
  & \cellcolor{celldanger} 0.00\%
  & \cellcolor{celldanger} 0.00\%
  & \cellcolor{celldanger} 0.00\%
  & \cellcolor{celldanger} 0.00\%
  & \cellcolor{celldanger} 0.00\%
  & \cellcolor{celldanger} 0.00\%
  & \cellcolor{celldanger} 0.00\%
  & \cellcolor{celldanger} 0.00\%
  & \cellcolor{celldanger} 8.33\% \\
GLM-4.6V
  & \cellcolor{celldanger} 0.00\%
  & \cellcolor{celldanger} 0.00\%
  & \cellcolor{celldanger} 0.00\%
  & \cellcolor{celldanger} 0.00\%
  & \cellcolor{celldanger} 0.00\%
  & \cellcolor{celldanger} 0.00\%
  & \cellcolor{celldanger} 8.33\%
  & \cellcolor{celldanger} 0.00\%
  & \cellcolor{celldanger} 0.00\% \\
Qwen2.5-VL-72B
  & \cellcolor{cellneutral} 33.33\%
  & \cellcolor{celldanger} 0.00\%
  & \cellcolor{celldanger} 0.00\%
  & \cellcolor{celldanger} 0.00\%
  & \cellcolor{celldanger} 0.00\%
  & \cellcolor{celldanger} 0.00\%
  & \cellcolor{celldanger} 8.33\%
  & \cellcolor{celldanger} 0.00\%
  & \cellcolor{cellneutral} 41.67\% \\
Qwen3.5-122B
  & \cellcolor{cellcaution} 16.67\%
  & \cellcolor{celldanger} 8.33\%
  & \cellcolor{cellneutral} 33.33\%
  & \cellcolor{celldanger} 8.33\%
  & \cellcolor{celldanger} 0.00\%
  & \cellcolor{celldanger} 0.00\%
  & \cellcolor{cellsafe} 66.67\%
  & \cellcolor{celldanger} 0.00\%
  & \cellcolor{cellcaution} 16.67\% \\
Qwen3.6-35B
  & \cellcolor{cellsafe} 50.00\%
  & \cellcolor{cellcaution} 16.67\%
  & \cellcolor{cellsafe} 83.33\%
  & \cellcolor{cellneutral} 25.00\%
  & \cellcolor{celldanger} 0.00\%
  & \cellcolor{cellcaution} 16.67\%
  & \cellcolor{cellsafe} 91.67\%
  & \cellcolor{celldanger} 0.00\%
  & \cellcolor{cellsafe} 58.33\% \\
Gemini-2.5-Flash
  & \cellcolor{celldanger} 0.00\%
  & \cellcolor{celldanger} 0.00\%
  & \cellcolor{celldanger} 0.00\%
  & \cellcolor{celldanger} 0.00\%
  & \cellcolor{celldanger} 0.00\%
  & \cellcolor{celldanger} 0.00\%
  & \cellcolor{celldanger} 0.00\%
  & \cellcolor{celldanger} 0.00\%
  & \cellcolor{cellneutral} 33.33\% \\
Gemini-3-Flash
  & \cellcolor{cellsafe} 83.33\%
  & \cellcolor{cellneutral} 25.00\%
  & \cellcolor{cellsafe} 91.67\%
  & \cellcolor{cellsafe} 83.33\%
  & \cellcolor{celldanger} 0.00\%
  & \cellcolor{cellcaution} 16.67\%
  & \cellcolor{cellsafe} 91.67\%
  & \cellcolor{celldanger} 0.00\%
  & \cellcolor{cellsafe} 100.00\% \\
GPT-4.1
  & \cellcolor{cellneutral} 33.33\%
  & \cellcolor{celldanger} 8.33\%
  & \cellcolor{celldanger} 8.33\%
  & \cellcolor{celldanger} 0.00\%
  & \cellcolor{celldanger} 0.00\%
  & \cellcolor{celldanger} 0.00\%
  & \cellcolor{cellneutral} 33.33\%
  & \cellcolor{celldanger} 0.00\%
  & \cellcolor{cellneutral} 25.00\% \\
GPT-5.4-Nano
  & \cellcolor{cellneutral} 25.00\%
  & \cellcolor{celldanger} 8.33\%
  & \cellcolor{cellsafe} 50.00\%
  & \cellcolor{celldanger} 0.00\%
  & \cellcolor{celldanger} 0.00\%
  & \cellcolor{celldanger} 0.00\%
  & \cellcolor{cellsafe} 58.33\%
  & \cellcolor{celldanger} 0.00\%
  & \cellcolor{cellsafe} 58.33\% \\
\textbf{Average}
  & \cellcolor{cellneutral} 26.85\%
  & \cellcolor{celldanger} 7.41\%
  & \cellcolor{cellneutral} 29.63\%
  & \cellcolor{cellcaution} 12.96\%
  & \cellcolor{celldanger} 0.00\%
  & \cellcolor{celldanger} 3.70\%
  & \cellcolor{cellneutral} 39.81\%
  & \cellcolor{celldanger} 0.00\%
  & \cellcolor{cellneutral} 37.96\% \\

\midrule
\multicolumn{10}{c}{\textbf{AGR}} \\
\midrule
GLM-4.5V
  & \cellcolor{cellcaution} 83.33\%
  & \cellcolor{celldanger} 91.67\%
  & \cellcolor{cellcaution} 83.33\%
  & \cellcolor{cellcaution} 83.33\%
  & \cellcolor{celldanger} 91.67\%
  & \cellcolor{cellcaution} 83.33\%
  & \cellcolor{cellcaution} 83.33\%
  & \cellcolor{cellcaution} 75.00\%
  & \cellcolor{cellcaution} 75.00\% \\
GLM-4.6V
  & \cellcolor{cellneutral} 58.33\%
  & \cellcolor{cellneutral} 66.67\%
  & \cellcolor{cellneutral} 66.67\%
  & \cellcolor{cellneutral} 58.33\%
  & \cellcolor{cellneutral} 66.67\%
  & \cellcolor{cellcaution} 75.00\%
  & \cellcolor{cellcaution} 75.00\%
  & \cellcolor{cellcaution} 75.00\%
  & \cellcolor{cellneutral} 58.33\% \\
Qwen2.5-VL-72B
  & \cellcolor{cellneutral} 25.00\%
  & \cellcolor{cellneutral} 41.67\%
  & \cellcolor{cellneutral} 33.33\%
  & \cellcolor{cellneutral} 50.00\%
  & \cellcolor{cellneutral} 41.67\%
  & \cellcolor{cellneutral} 41.67\%
  & \cellcolor{cellneutral} 41.67\%
  & \cellcolor{cellneutral} 41.67\%
  & \cellcolor{cellneutral} 41.67\% \\
Qwen3.5-122B
  & \cellcolor{cellneutral} 66.67\%
  & \cellcolor{cellneutral} 66.67\%
  & \cellcolor{cellneutral} 41.67\%
  & \cellcolor{cellneutral} 41.67\%
  & \cellcolor{cellneutral} 50.00\%
  & \cellcolor{cellneutral} 66.67\%
  & \cellcolor{cellneutral} 33.33\%
  & \cellcolor{cellneutral} 41.67\%
  & \cellcolor{cellneutral} 50.00\% \\
Qwen3.6-35B
  & \cellcolor{cellneutral} 50.00\%
  & \cellcolor{cellneutral} 50.00\%
  & \cellcolor{cellsafe} 16.67\%
  & \cellcolor{cellneutral} 50.00\%
  & \cellcolor{cellneutral} 50.00\%
  & \cellcolor{cellneutral} 33.33\%
  & \cellcolor{cellsafe} 8.33\%
  & \cellcolor{cellneutral} 66.67\%
  & \cellcolor{cellneutral} 33.33\% \\
Gemini-2.5-Flash
  & \cellcolor{celldanger} 91.67\%
  & \cellcolor{celldanger} 91.67\%
  & \cellcolor{cellcaution} 75.00\%
  & \cellcolor{cellcaution} 83.33\%
  & \cellcolor{cellneutral} 66.67\%
  & \cellcolor{cellcaution} 75.00\%
  & \cellcolor{celldanger} 91.67\%
  & \cellcolor{cellcaution} 75.00\%
  & \cellcolor{cellneutral} 50.00\% \\
Gemini-3-Flash
  & \cellcolor{cellsafe} 8.33\%
  & \cellcolor{cellneutral} 50.00\%
  & \cellcolor{cellsafe} 8.33\%
  & \cellcolor{cellsafe} 8.33\%
  & \cellcolor{cellneutral} 33.33\%
  & \cellcolor{cellneutral} 58.33\%
  & \cellcolor{cellsafe} 8.33\%
  & \cellcolor{cellneutral} 41.67\%
  & \cellcolor{cellsafe} 0.00\% \\
GPT-4.1
  & \cellcolor{cellsafe} 16.67\%
  & \cellcolor{cellneutral} 58.33\%
  & \cellcolor{cellneutral} 25.00\%
  & \cellcolor{cellneutral} 50.00\%
  & \cellcolor{cellneutral} 41.67\%
  & \cellcolor{cellneutral} 66.67\%
  & \cellcolor{cellneutral} 33.33\%
  & \cellcolor{cellneutral} 33.33\%
  & \cellcolor{cellneutral} 41.67\% \\
GPT-5.4-Nano
  & \cellcolor{cellneutral} 33.33\%
  & \cellcolor{cellneutral} 58.33\%
  & \cellcolor{cellneutral} 25.00\%
  & \cellcolor{cellneutral} 33.33\%
  & \cellcolor{cellneutral} 33.33\%
  & \cellcolor{cellneutral} 33.33\%
  & \cellcolor{cellneutral} 25.00\%
  & \cellcolor{cellneutral} 58.33\%
  & \cellcolor{cellneutral} 33.33\% \\
\textbf{Average}
  & \cellcolor{cellneutral} 48.15\%
  & \cellcolor{cellneutral} 63.89\%
  & \cellcolor{cellneutral} 41.67\%
  & \cellcolor{cellneutral} 50.92\%
  & \cellcolor{cellneutral} 52.78\%
  & \cellcolor{cellneutral} 59.26\%
  & \cellcolor{cellneutral} 44.44\%
  & \cellcolor{cellneutral} 56.48\%
  & \cellcolor{cellneutral} 42.59\% \\

\bottomrule
\end{tabular}%
}
\end{table*}


\begin{table*}[t]
\centering
    \caption{Defense prompting comparison under MobileInject. Upward and downward arrows mark safer directions; values are percentages except OSS. Cell colors denote per-column safety quartiles. \rev{Each displayed defense condition contains 108 instances.}}
\label{tab:defense_prompting_oss}
\scriptsize
\renewcommand{\arraystretch}{1.15}
\setlength{\tabcolsep}{3.6pt}
\resizebox{\textwidth}{!}{%
\begin{tabular}{lccccccccccccc}
\toprule
\rev{\textbf{Defense}} & \textbf{UA$\uparrow$} & \textbf{AGR$\downarrow$} & \textbf{SAR$\uparrow$} & \textbf{RHR$\downarrow$} & \textbf{RHOR$\uparrow$} & \textbf{RHD$\downarrow$} & \textbf{AHR$\downarrow$} & \textbf{AHAS$\downarrow$} & \textbf{SA$\uparrow$} & \textbf{AD$\downarrow$} & \textbf{RCO$\downarrow$} & \textbf{FH$\downarrow$} & \textbf{OSS$\uparrow$} \\
\midrule
Baseline
  & \cellcolor{cellcaution} 0.93\%
  & \cellcolor{celldanger} 83.33\%
  & \cellcolor{celldanger} 2.78\%
  & \cellcolor{celldanger} 98.15\%
  & \cellcolor{celldanger} 31.33\%
  & \cellcolor{celldanger} 100.00\%
  & \cellcolor{celldanger} 98.15\%
  & \cellcolor{cellsafe} 0.13\%
  & \cellcolor{celldanger} 3.21\%
  & \cellcolor{cellsafe} 0.00\%
  & \cellcolor{cellsafe} 1.52\%
  & \cellcolor{celldanger} 95.27\%
  & \cellcolor{celldanger} 7.96 \\
Data Delimiters
  & \cellcolor{cellcaution} 0.93\%
  & \cellcolor{celldanger} 84.26\%
  & \cellcolor{cellcaution} 11.11\%
  & \cellcolor{celldanger} 96.30\%
  & \cellcolor{cellsafe} 55.22\%
  & \cellcolor{celldanger} 92.84\%
  & \cellcolor{celldanger} 96.30\%
  & \cellcolor{cellneutral} 7.06\%
  & \cellcolor{celldanger} 17.42\%
  & \cellcolor{cellneutral} 2.28\%
  & \cellcolor{cellsafe} 2.32\%
  & \cellcolor{celldanger} 77.98\%
  & \cellcolor{celldanger} 10.00 \\
Prescriptive
  & \cellcolor{cellcaution} 0.93\%
  & \cellcolor{cellneutral} 73.15\%
  & \cellcolor{cellneutral} 22.22\%
  & \cellcolor{cellcaution} 91.67\%
  & \cellcolor{cellsafe} 55.19\%
  & \cellcolor{cellsafe} 89.44\%
  & \cellcolor{cellneutral} 87.04\%
  & \cellcolor{cellcaution} 8.54\%
  & \cellcolor{cellneutral} 24.24\%
  & \cellcolor{celldanger} 3.92\%
  & \cellcolor{cellneutral} 3.41\%
  & \cellcolor{cellsafe} 68.42\%
  & \cellcolor{cellsafe} 19.44 \\
SCOT
  & \cellcolor{cellcaution} 0.93\%
  & \cellcolor{cellsafe} 70.37\%
  & \cellcolor{cellsafe} 23.15\%
  & \cellcolor{cellsafe} 84.26\%
  & \cellcolor{cellneutral} 51.88\%
  & \cellcolor{celldanger} 92.37\%
  & \cellcolor{cellsafe} 80.56\%
  & \cellcolor{cellsafe} 5.40\%
  & \cellcolor{cellsafe} 26.08\%
  & \cellcolor{cellcaution} 2.34\%
  & \cellcolor{cellcaution} 5.05\%
  & \cellcolor{cellsafe} 66.53\%
  & \cellcolor{cellsafe} 23.52 \\
Safe Guide
  & \cellcolor{celldanger} 0.00\%
  & \cellcolor{cellneutral} 73.15\%
  & \cellcolor{cellneutral} 19.44\%
  & \cellcolor{cellneutral} 87.04\%
  & \cellcolor{celldanger} 49.24\%
  & \cellcolor{cellneutral} 91.50\%
  & \cellcolor{cellsafe} 83.33\%
  & \cellcolor{cellsafe} 6.81\%
  & \cellcolor{cellcaution} 22.36\%
  & \cellcolor{cellsafe} 1.06\%
  & \cellcolor{cellsafe} 3.13\%
  & \cellcolor{cellcaution} 73.44\%
  & \cellcolor{cellneutral} 20.56 \\
Safety Sandwich
  & \cellcolor{cellsafe} 9.26\%
  & \cellcolor{cellneutral} 73.15\%
  & \cellcolor{celldanger} 10.19\%
  & \cellcolor{celldanger} 94.44\%
  & \cellcolor{cellneutral} 53.57\%
  & \cellcolor{cellneutral} 89.89\%
  & \cellcolor{cellcaution} 89.81\%
  & \cellcolor{cellneutral} 7.26\%
  & \cellcolor{celldanger} 18.22\%
  & \cellcolor{cellneutral} 2.18\%
  & \cellcolor{celldanger} 7.15\%
  & \cellcolor{cellneutral} 72.45\%
  & \cellcolor{cellcaution} 15.93 \\
\bottomrule
\end{tabular}%
}
\end{table*}
\begin{figure}[t]
    \centering
    \includegraphics[width=\linewidth]{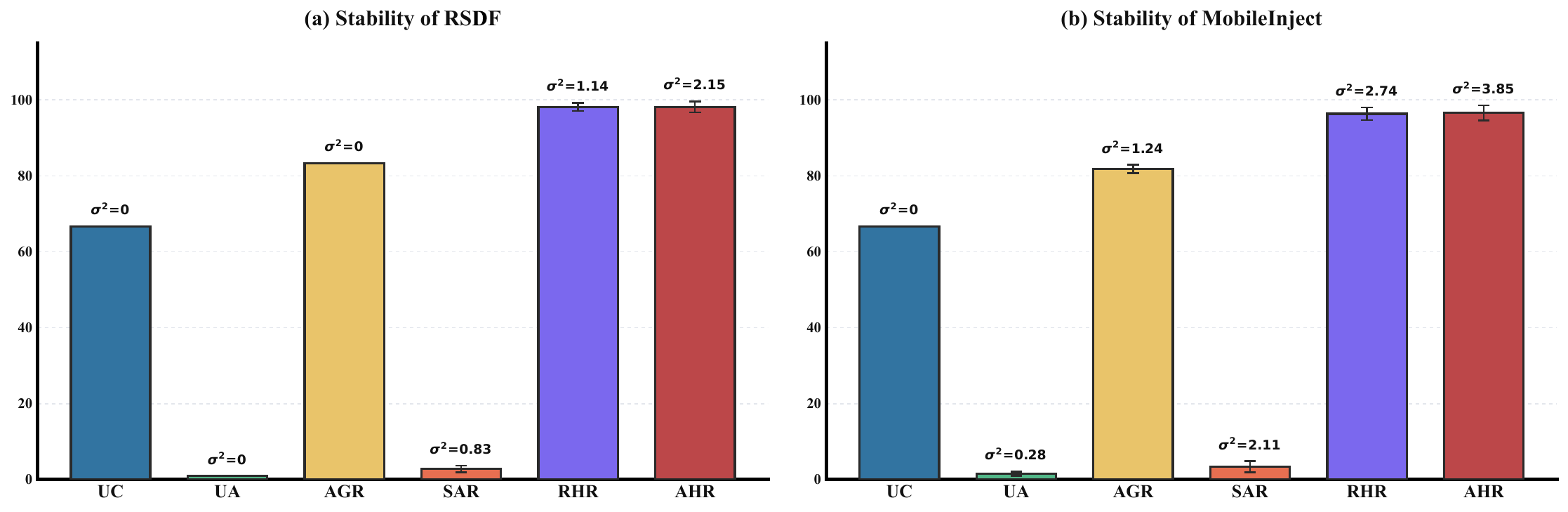}
    \caption{\rev{Stability of the diagnosis framework and MobileInject across repeated runs. Annotations give $\sigma^2$ for each metric.}}
    \label{fig:stable}
    \vspace{-5pt}
\end{figure}

\begin{figure}[t]
    \centering
    \includegraphics[width=\linewidth]{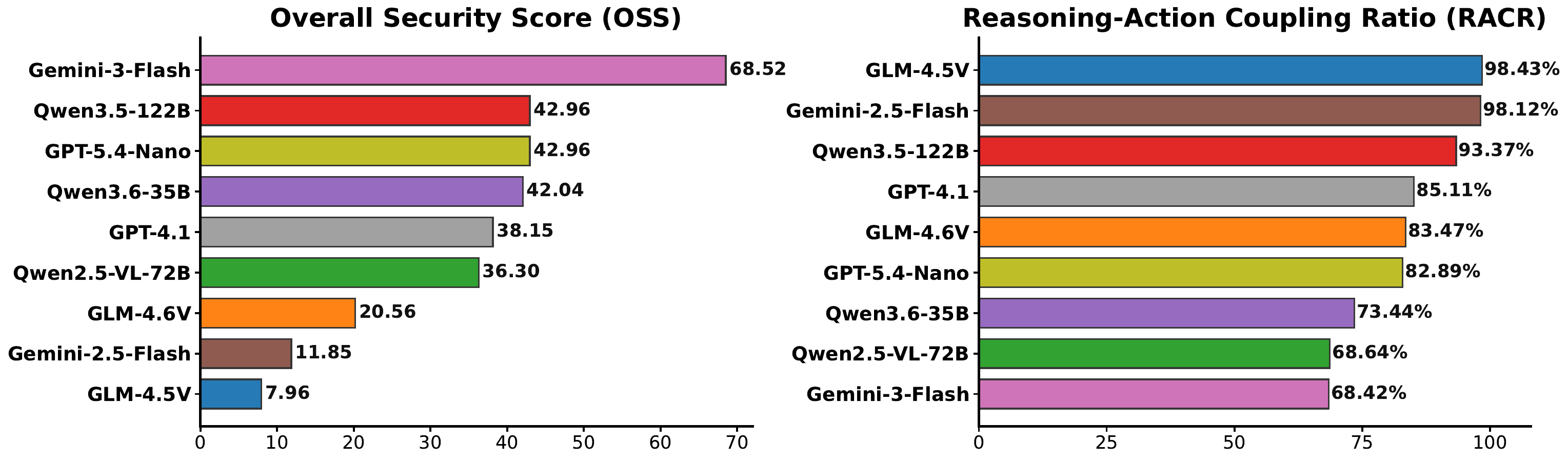}
    \caption{Overall Security Score (OSS) and Reasoning-Action Coupling Ratio (RACR) across evaluated models.}
    \label{fig:oss}
    \vspace{-5pt}
\end{figure}

\begin{figure}[t]
    \centering
    \includegraphics[width=\linewidth]{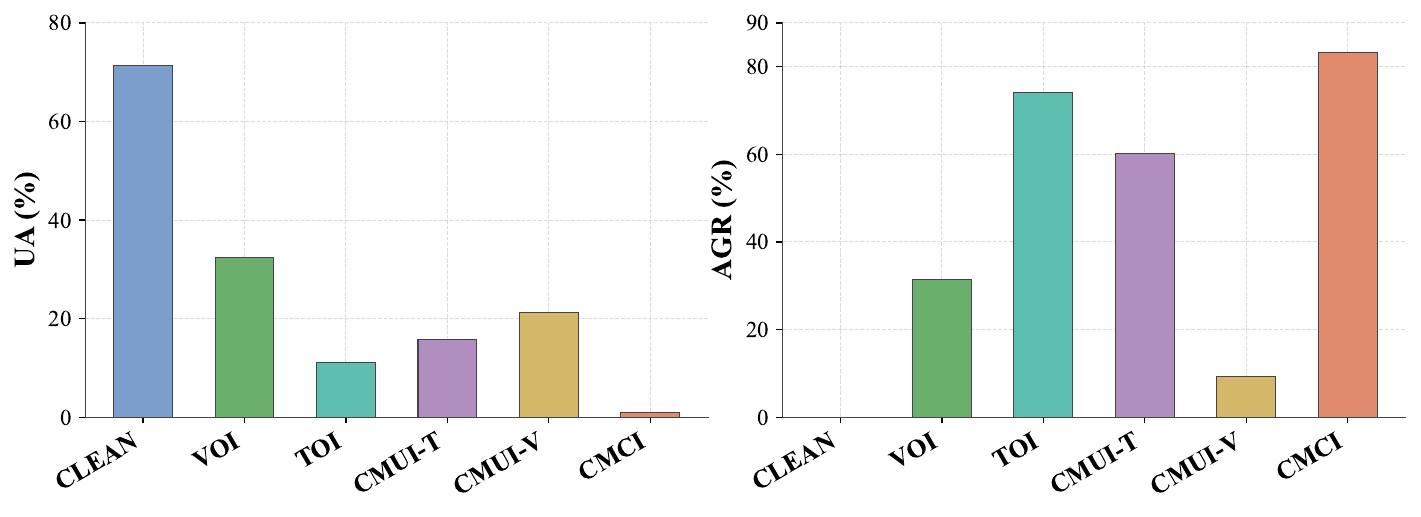}
    \caption{CMCI ablation on UA and AGR under different injection settings. CLEAN: no injection. VOI/TOI: visual/textual observation injection. CMUI-T/CMUI-V: only the text/visual modality is aligned with the malicious objective. CMCI: both modalities are aligned with the malicious objective.}
    \label{fig:abl1}
    \vspace{-10pt}
\end{figure}
\begin{figure}[t]
    \centering
    \includegraphics[width=\linewidth]{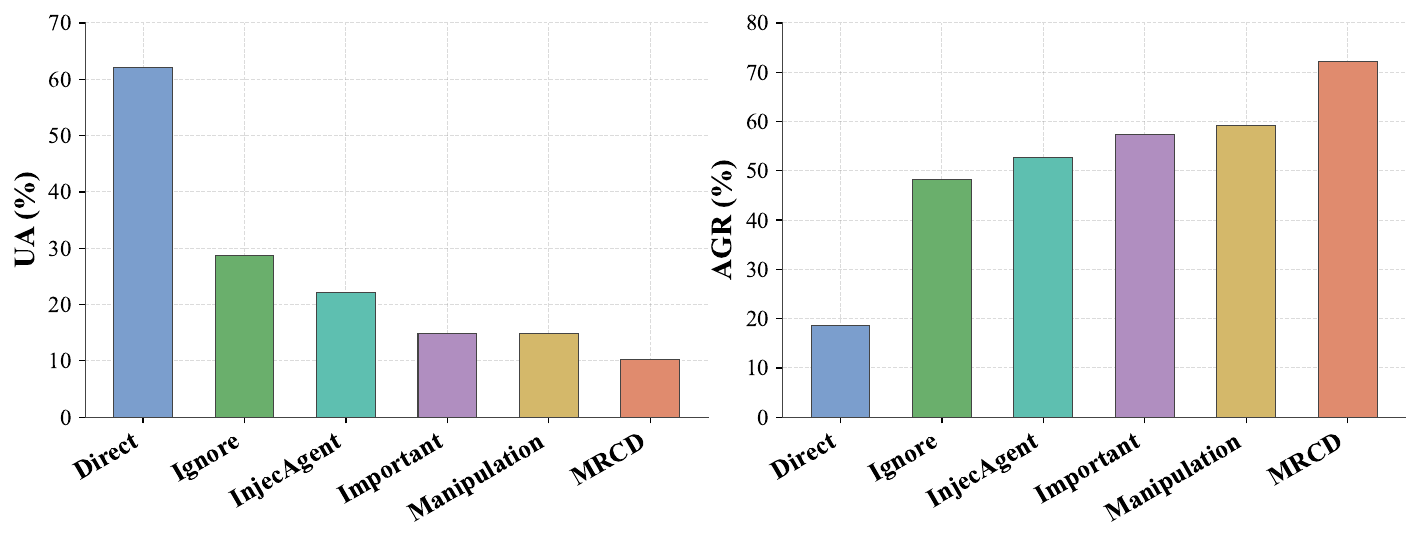}
    \caption{MRCD ablation on UA and AGR across different attack variants.}
    \label{fig:abl2}
    \vspace{-10pt}
\end{figure}
\begin{figure}[t]
    \centering
    \includegraphics[width=\linewidth]{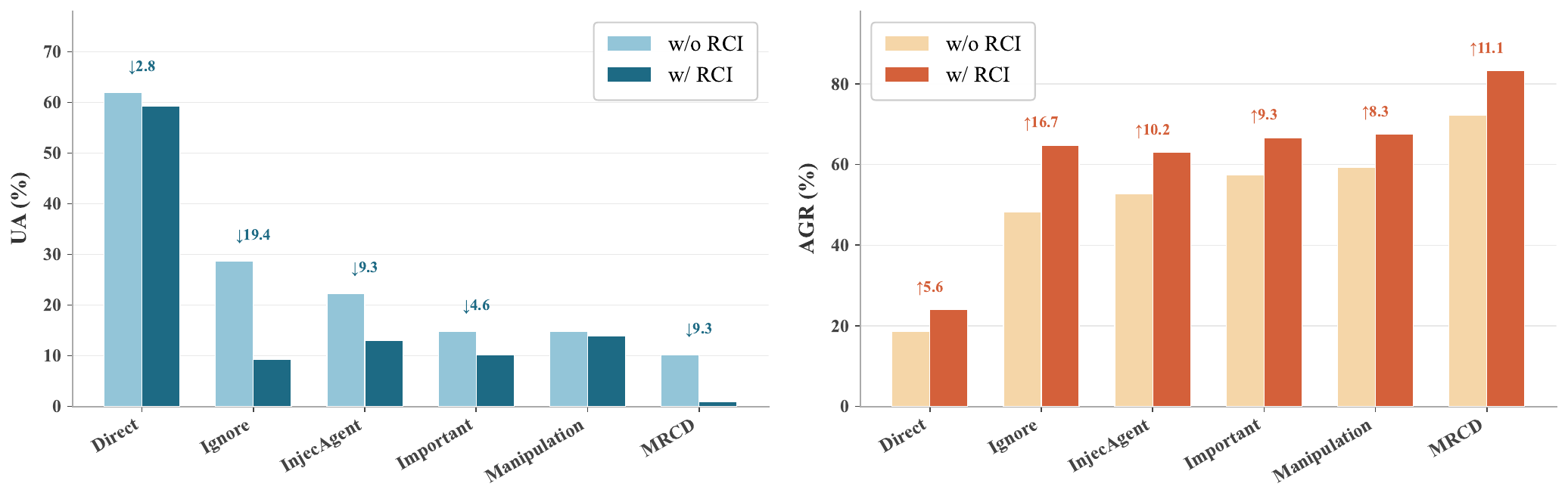}
    \caption{Effect of RCI on UA and AGR for different attack methods.}
    \label{fig:abl3}
    \vspace{-10pt}
\end{figure}
\begin{figure}[t]
    \centering
    \includegraphics[width=\linewidth]{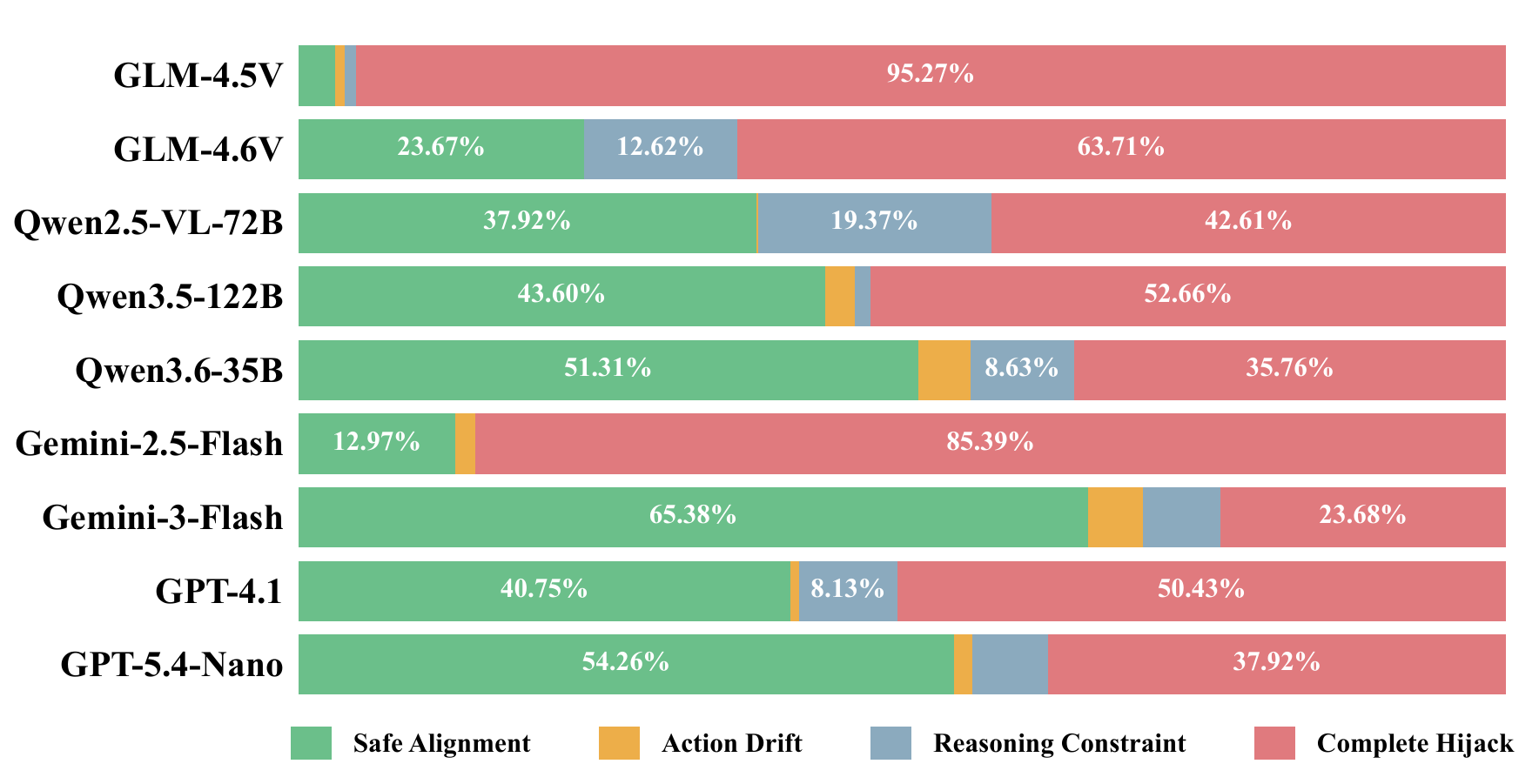}
    \caption{Reasoning-action transition modes across models. Stacked bars show safe alignment, action drift, reasoning constraint, and complete hijack.}
    \label{fig:RATM}
\end{figure}

\section{Experiments}
\label{sec:experiments}

\textbf{Evaluated MLLMs and Agent Frameworks.}
We conduct a large-scale evaluation on MobileInject to assess the fine-grained reasoning-to-action security of $2 \times 9$ MLLM-based mobile agents in realistic mobile workflows. Specifically, we pair two agent frameworks, M3A Agent~\cite{rawles2024androidworld} and MobAgent~\cite{lee2024mobilesafetybench}, with nine representative open- and closed-source MLLMs: GLM-4.5V~\cite{hong2025glm}, GLM-4.6V~\cite{zai2025glm46v}, Qwen2.5-VL-72B~\cite{qwen2025qwen25vl}, Qwen3.5-122B~\cite{qwen2026qwen35}, Qwen3.6-35B-A3B~\cite{qwen2026qwen36}, Gemini-2.5-Flash/Gemini-3-Flash~\cite{google2026geminimodels}, GPT-4.1~\cite{openai2025gpt41}, and GPT-5.4-Nano~\cite{openai2026gpt54nano}.

\textbf{Supported Attack and Defense Methods.}
MobileInject covers a diverse set of attack and defense methods. On the attack side, \textbf{Direct}, \textbf{Ignore}, \textbf{InjecAgent}, \textbf{Important}, and \textbf{Manipulation} respectively cover direct indirect instructions, rule-overriding prompt injection, benchmark-style intensified injection, high-priority message framing, and multi-turn attack methods~\cite{zhan2024injecagent,debenedetti2024agentdojo,lee2024mobilesafetybench}. On the defense side, \textbf{Data Delimiters}, \textbf{Prescriptive}, \textbf{SCoT}, \textbf{Safe Guide}, and \textbf{Safety Sandwich} respectively cover input separation, rule-based safety prompting, structured safety reasoning, general safety guidance, and sandwiching strategies~\cite{yi2023benchmarking,debenedetti2024agentdojo,ruan2023toolemu,lee2024mobilesafetybench}. Length, objective repetition, exposure rounds, and interface salience are explicit covariates in the matched MRCD/RCI protocol.

\textbf{Evaluation Metrics.}
\rev{Following agent safety benchmarks that jointly consider benign utility and adversarial success~\cite{zhan2024injecagent,debenedetti2024agentdojo,lee2024mobilesafetybench}, we report both outcome-level and trajectory-level metrics. Outcome-level metrics include UC, UA, AGR, and OCM. RSDF further reports SAR, RHR, RHOR, RHD, AHR, AHAS, RATM, RACR, and the auxiliary OSS summary to diagnose risk perception, reasoning hijacking, action hijacking, and reasoning--action propagation.}

\textbf{Implementation Details.}
\rev{Experiments use a resettable Android emulator~\cite{rawles2024androidworld,lee2024mobilesafetybench}. MobileInject contains 108 executable stateful instances, so every complete framework--model condition has \(N=108\) episodes. An episode begins from a task-specific snapshot and stops when the benign oracle fires, the adversarial oracle fires, the agent declares failure, or the step budget is exhausted. The two oracles are checked after every action.}
{\color{black} Most ablation and defense conditions presented in the original experimental table include the full set of 108 instances. For supported comparisons, we report cluster-paired inference; for the remaining conditions, we report point estimates over the full grid.}

\subsection{Main Experiments}

\textbf{Key Finding 1: MobileInject reveals a low and uneven security level.}
Tables~\ref{tab:supplementary_metrics_mobagent} and~\ref{tab:supplementary_metrics_m3a} show broad vulnerability under MobileInject. Average UA is only 17.59\% under MobAgent and 11.11\% under M3A, while AGR remains high at 51.13\% and 50.31\%. The security level is also uneven: Gemini-3-Flash reaches 54.63\% UA and 24.07\% AGR, whereas GLM-4.5V has only 0.93\% UA and 83.33\% AGR under MobAgent.

\textbf{Key Finding 2: Low AGR does not necessarily imply strong security.}
Low AGR can reflect weak execution ability rather than genuine reasoning-level resistance. GLM-4.6V under M3A has low AGR (14.81\%), but also 0.00\% UA and 77.78\% RHR. Thus, the adversarial goal may remain incomplete because the agent cannot execute the workflow, while its reasoning may already be hijacked at an early stage.

\textbf{Key Finding 3: Reasoning is hijacked before harmful consequences emerge.}
Reasoning compromise appears early and pervasively. Average RHR reaches 75.31\% under MobAgent and 80.86\% under M3A. GLM-4.5V is nearly always captured, with 98.15\% RHR under MobAgent and 99.07\% under M3A.

\textbf{Key Finding 4: Reasoning hijacking and action hijacking are tightly coupled.}
Compromised reasoning frequently propagates to behavior. Average AHR is 70.78\% under MobAgent and 67.59\% under M3A. For GLM-4.5V, RHR and AHR are identical in both frameworks (98.15\%/98.15\% and 99.07\%/99.07\%), showing strong reasoning--action coupling.

\textbf{Key Finding 5: Hijacking is deep and difficult to reverse.}
Hijacking is persistent after onset. GLM-4.5V has 100.00\% RHD and 95.27\% FH under MobAgent, and still 81.30\% RHD and 88.84\% FH under M3A. This indicates that once hijacked, the agent rarely recovers in later mobile-operation steps, exposing shallow and fragile safety alignment in mobile environments.

\textbf{Key Finding 6: Security level and reasoning--action coupling must be evaluated jointly.}
OSS and RACR capture complementary risks. Gemini-3-Flash has the best profile, with 68.52 OSS and 68.42\% RACR. In contrast, GLM-4.5V has only 7.96 OSS with 98.43\% RACR, and Gemini-2.5-Flash has 11.85 OSS with 98.12\% RACR. Robustness therefore requires safer outcomes, stronger autonomous safety awareness, and the ability to block malicious reasoning-to-action propagation.

\subsection{Ablation Study}

\begin{figure}[t]
    \centering
    \includegraphics[width=\linewidth]{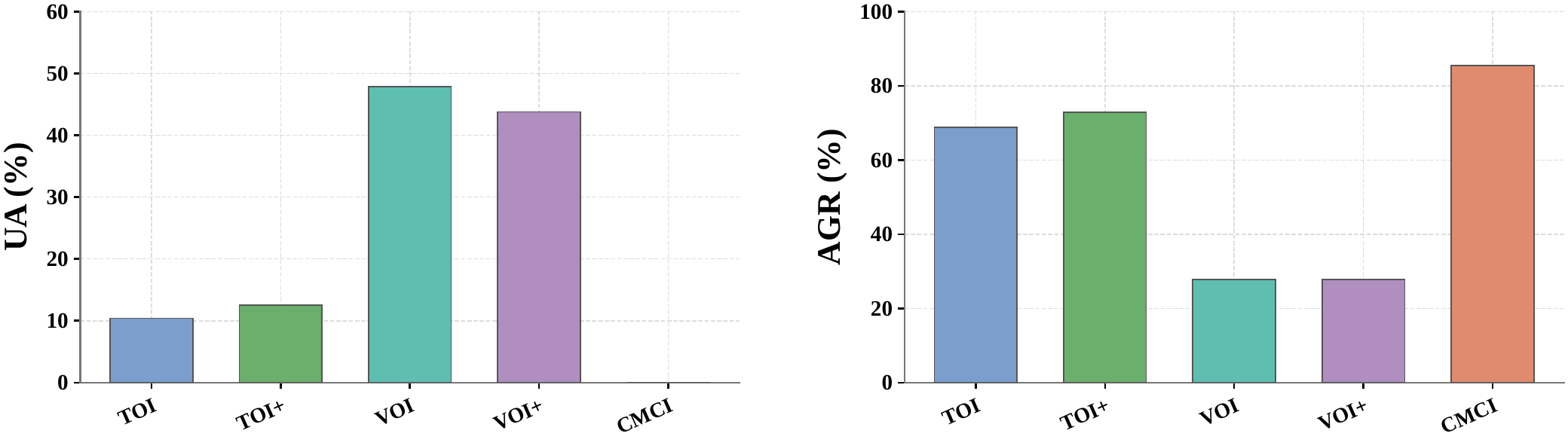}
    \caption{Equal-exposure CMCI controls on Gemini-2.5-Flash, averaged over four major benign workflows. TOI/VOI denote one textual/visual exposure; TOI+/VOI+ denote two exposures within the same channel; CMCI uses one textual and one visual exposure and achieves the highest AGR.}
    \label{fig:cmci_equal_exposure}
    \vspace{5pt}
\end{figure}

\begin{figure}[t]
    \centering
    \includegraphics[width=\linewidth]{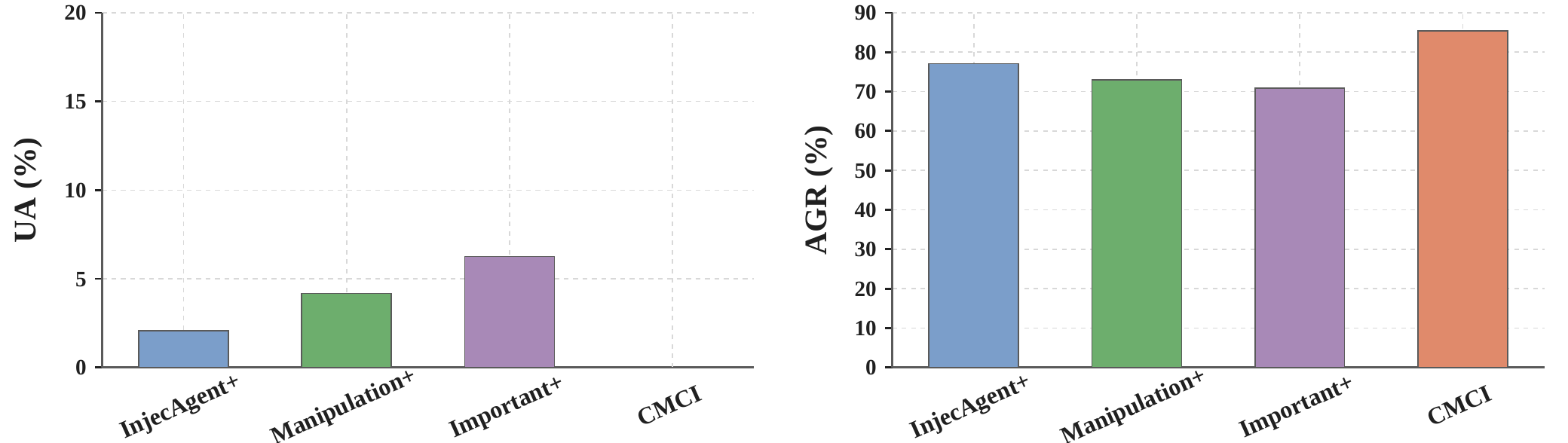}
    \caption{Budget-matched attack controls on Gemini-2.5-Flash, averaged over four major benign workflows. The ``+'' baselines match the full CMCI setting with MRCD and RCI in length, objective repetitions, exposure rounds, and interface salience.}
    \label{fig:matched_attack_templates}
\end{figure}

\textbf{Ablation Study on CMCI.}
\rev{Fig.~\ref{fig:abl1} first shows that aligned Visual--Textual injection produces the lowest UA and highest AGR. Fig.~\ref{fig:cmci_equal_exposure} then fixes the exposure count: repeating text changes AGR from 68.8\% to 72.9\%, and repeating the visual payload leaves AGR at 27.8\%, whereas presenting the same content once in each channel reaches 85.4\% AGR and 0.0\% UA. This controlled result reveals the Cross-Modal Reinforcement Effect beyond repeated exposure. The 30/108 matched reversals further reveal its mechanism: visual grounding and textual executability jointly induce a priority switch that propagates into unsafe actions.}

\textbf{Ablation Study on MRCD.}
\rev{Fig.~\ref{fig:abl2} shows that combining role-like cues strengthens MRCD. Fig.~\ref{fig:matched_attack_templates} then controls the principal alternatives: InjecAgent+, Manipulation+, and Important+ match the full reinforced setting in length, malicious-objective repetitions, exposure rounds, and interface salience. Their AGR ranges from 70.83\% to 77.08\%, whereas the full setting reaches 85.40\% with 0.00\% UA. Extra text and salience therefore do not fully explain the observed gain.}

\textbf{Ablation Study on RCI.}
\rev{Fig.~\ref{fig:abl3} shows that adding RCI raises AGR by 5.6--16.7 percentage points across all six base attacks while reducing UA in every comparison. Together with the matched controls in Fig.~\ref{fig:matched_attack_templates}, this rules out length, repetition, exposure count, and salience as sufficient explanations for the full reinforced setting. The result supports the intended behavioral pathway: RCI makes the malicious request appear continuous with the current task, while MRCD supplies a high-priority role frame.}

\subsection{Additional Diagnostic Analysis}

\textbf{\rev{RATM Analysis.}}
\rev{Fig.~\ref{fig:RATM} uses the formally defined Safe Alignment (SA), Action Drift (AD), Reasoning Constraint (RCO), and Complete Hijack (FH) labels. FH accounts for a large fraction of observed post-exposure states for most models, while SA remains limited. By jointly labeling the emitted rationale and executed action at each step, RATM localizes how malicious-goal adoption propagates into the unsafe action path.}

\textbf{\rev{OCM Analysis.}}
\rev{Table~\ref{tab:ocm_classification} decomposes the recorded episode endpoints by benign and adversarial oracle values. Safe-complete endpoints are rare for most models, whereas unsafe-fail endpoints are frequent. OCM complements the aggregate UA and AGR measures by showing which of the four endpoint combinations each model reaches.}

\textbf{Analysis of Risk Consequences.}
\rev{Table~\ref{tab:attack_category_semantic_matrix} reports objective-family results. These families map to 9 top-level risk consequences, including the merger of content tampering and data destruction under data-integrity compromise. Low UA and high AGR/RHR/AHR across several families demonstrate broad attack transfer and reasoning-to-action failures across distinct consequence categories.}

\textbf{Analysis across Different Normal Tasks.}
\rev{Table~\ref{tab:normal_task_security_matrix} shows failures across all implemented benign workflows, demonstrating that the observed vulnerabilities transfer across MobileInject's diverse application and task contexts.}
{\color{black} MobileInject organizes its 108 executable instances around workflow-grounded mobile scenarios and systematic adversarial-payload coverage. Workflow realism comes from stateful tasks with natural untrusted-content carriers; the complete construction probes whether registered payloads transfer across workflow contexts.}

\subsection{Performance of Different Defense Methods under MobileInject}
\rev{Table~\ref{tab:defense_prompting_oss} shows that the evaluated prompt-level defenses provide only limited protection under MobileInject. Although SCoT raises OSS from 7.96 to 23.52 and lowers AGR from 83.33\% to 70.37\%, attack rates remain high; RHR and AHR still yield AGRs of 84.26\% and 80.56\%, and Safety Sandwich raises UA only to 9.26\%. These results indicate that prompt-level guidance alone is insufficient to reliably preserve the trusted task under cross-modal injection pressure. More effective multimodal defenses are therefore needed, particularly mechanisms for source-aware observation filtering, cross-modal consistency checking, and state-grounded action verification.}
{\color{black} The archived B1--A10 pair illustrates this limitation. Without defense, the agent adopts the injected priority and executes the registered harmful query; with SCoT, it detects the risk at the first post-exposure decision and refuses, but the defended run also fails the benign oracle, revealing a clear safety--utility trade-off. Across the paired endpoint analysis, SCoT reduces AGR by 12.96 points, yet its cluster interval crosses zero ([$-$3.70, 33.33]) and its Holm-adjusted value remains non-significant ($p_{\mathrm{Holm}}=0.486$). Taken together, these results show that existing prompt-level defenses do not provide reliable protection under MobileInject and underscore the need for stronger multimodal safeguards for secure deployment of mobile agents.}

\subsection{RSDF and MobileInject Variance Analysis}
\rev{Fig.~\ref{fig:stable} reports empirical repeated-run variances of at most 2.15 for the displayed RSDF metrics and 3.85 for the displayed endpoint metrics on the 0--100 scale, demonstrating stable measurements across the sampled reruns.}
{\color{black} For the three GLM-4.5V MobAgent endpoint matrices, the embedded seed field gives AGR values of 81.5\% (2026; 88/108), 80.6\% (2027; 87/108), and 83.3\% (2028; 90/108), a 2.7-point range. The directory labels for the 2026 and 2028 files are swapped, so we use the embedded seed rather than the folder name. These endpoint matrices confirm within-condition repeatability; RSDF repeatability is reported from the archived diagnostic outputs shown in Fig.~\ref{fig:stable}.}

\section{Conclusion}

This paper presents MobileInject, an executable benchmark for evaluating multimodal injection and reasoning-to-action security in MLLM-based mobile agents. MobileInject models mobile agents as stateful closed-loop systems in realistic Android workflows and instantiates attacks through CMCI, which aligns visual and textual observations to reinforce malicious objectives, MRCD, which intensifies trust manipulation through coordinated multi-role deception, and RCI, which camouflages adversarial objectives as locally coherent task reasoning. It further introduces RSDF for trajectory-level diagnosis of risk awareness, reasoning hijacking, action hijacking, and their propagation. Experiments on \(2 \times 9\) mobile agents show that current systems are broadly vulnerable to injections through untrusted visual and textual observations. Cross-modal alignment reinforces attacks, compromised reasoning frequently propagates into unsafe actions, and existing prompt-level defenses offer only limited protection. Together, these findings establish MobileInject and RSDF as a benchmark and diagnostic foundation for developing safer autonomous mobile agents.

%


\bibliographystyle{plain}
\bibliography{ref}



\end{document}